\documentclass[12pt,preprint]{aastex}
\usepackage{epsfig}
\usepackage{psfig}

\newcommand{\chisq}{$\chi^2$}

\newcommand{\PSRB}{PSR B0950+08}
\newcommand{\PSRJ}{PSR J2043+2740}
\newcommand{\PSRBB}{PSR B0823+26}

\newcommand \delp {\Delta^{\rm pk}}

\shorttitle{{XMM-Newton Observations of PSR B0950+08, B0823+26 and J2043+2740}}
\shortauthors{W.~Becker et al.}
\begin{document}

\title{{Revealing the X-ray emission processes of old rotation-powered pulsars:
XMM-Newton Observations of PSR B0950+08,\newline PSR B0823+26 and PSR J2043+2740}}

\author{
Werner Becker\altaffilmark{1},
Martin C. Weisskopf\altaffilmark{2},
Allyn F. Tennant\altaffilmark{2},
Axel Jessner\altaffilmark{3},\\
Jaros{\l}aw Dyks\altaffilmark{4,}\altaffilmark{6},
Alice K. Harding\altaffilmark{4},
and Shuang N. Zhang\altaffilmark{2,}\altaffilmark{5} }

\altaffiltext{1}
{Max-Planck Institut f\"ur Extraterrestrische Physik, 85741 Garching bei M\"unchen, Germany}
\altaffiltext{2}
{Space Sciences Department, NASA Marshall Space Flight Center, SD50, Huntsville, AL 35812}
\altaffiltext{3}
{Max-Planck Institut f\"ur Radioastronomie, Effelsberg, 53902 Bad M\"unstereifel, Germany}
\altaffiltext{4}
{Laboratory for High-Energy Astrophysics, NASA Goddard Space Flight Center, Greenbelt, MD 20771}
\altaffiltext{5}
{Physics Department, University of Alabama in Huntsville, Huntsville, AL 35899}
\altaffiltext{6}
{On leave from Nicolaus Copernicus Astronomical Center, Toru\'n, Poland}

\begin{abstract}

We have completed part of a program to study the X-ray emission properties of
old rotation-powered pulsars with XMM-Newton in order to probe and identify
the origin of their X-radiation. The X-ray emission from these old pulsars is
largely dominated by non-thermal processes. None of the observed spectra
required adding a thermal component consisting of either a hot polar cap
or surface cooling emission to model the data.
The X-ray spectrum of \PSRB\, is best described by a single power law of
photon-index $\alpha=1.93^{+0.14}_{-0.12}$. Taking optical data from the 
VLT FORS1 into account a broken power law model with the break 
point $E_{break}= 0.67^{+0.18}_{-0.41}$ keV and the photon-index 
$\alpha_1= 1.27^{+0.02}_{-0.01}$ and $\alpha_2 =1.88^{+0.14}_{-0.11}$
for $E < E_{break}$ and $E > E_{break}$, respectively, is found to describe
the pulsar's broadband spectrum from the optical to the X-ray band. 
Three-$\sigma$ temperature upper limits for 
possible contributions from a heated polar cap or the whole neutron star 
surface are $T^\infty_{pc} < 0.87 \times 10^6$ K and $T^\infty_s < 0.48
\times 10^6$ K, respectively. We also find that the X-ray emission from \PSRB\,
is pulsed with two peaks per rotation period. The phase separation between the
two X-ray peaks is $\sim 144^\circ$ (maximum to maximum) which is similar to the
pulse peak separation observed in the radio band at 1.4 GHz. The main radio peak 
and the trailing X-ray peak are almost phase aligned. The fraction of X-ray 
pulsed photons is $\sim 30\%$. A phase-resolved spectral analysis confirms the 
non-thermal nature of the pulsed emission and finds no spectral variations as a
function of pulse phase.
Detailed pulse profile simulations using polar gap, outer gap and the two-pole 
caustic model constrain the pulsar's emission geometry to be that of an almost 
orthogonal rotator, for which the two-pole caustic model can reproduce the 
observed doubly peaked X-ray pulse profile. 
The spectral emission properties observed for \PSRBB\, are similar to those of \PSRB.
Its energy spectrum is very well described by a single power law with photon-index
$\alpha=2.5^{+0.9}_{-0.45}$. Three-$\sigma$ temperature upper limits for thermal
contributions from a hot polar cap or from the entire neutron star surface are
$T^\infty_{pc} <1.17 \times 10^6$ K and $T^\infty_s < 0.5 \times 10^6$ K,
respectively. There is evidence for pulsed X-ray emission at the
$\sim 97\%$ confidence level with a pulsed fraction of $49 \pm 22\%$.
For \PSRJ\, which is located $\sim 1^\circ$ outside the boundary of the Cygnus-Loop,
we report the first detection of X-ray emission. A power law spectrum, or a 
combination of a thermal and a power law spectrum all yield acceptable descriptions 
of its X-ray spectrum. No X-ray pulses are detected from \PSRJ\, and the sensitivity 
is low - the $2\sigma$ pulsed fraction upper limit is $57\%$ assuming a sinusoidal 
pulse profile.

\end{abstract}


\section{INTRODUCTION \label{intro}}

In the past decade, many advances have been made in the study of rotation
powered pulsars thanks to X- and $\gamma$-ray observations with a series of
high energy astrophysical space missions (for reviews see e.g.~Becker \&
Tr\"umper 1997; Becker \& Pavlov 2001; and Kanbach 2001). However, many outstanding
scientific issues still remain to be resolved. Understanding the high energy
emission processes of old rotation-powered pulsars is one of them.

Although one could use the instruments aboard ROSAT, BeppoSAX and ASCA to
disentangle the thermal and non-thermal contributions of the young and
cooling neutron stars, the sensitivity of these instruments was not
sufficient to study the emission properties of the old field pulsars.

Old rotation-powered field pulsars are of particular interest for the study
of particle acceleration and high energy radiation processes on the neutron
star surface and in the neutron star magnetosphere. This is because their ages
are intermediate between that of the young cooling neutron stars, whose
surface may produce copious thermal X-ray photons, and those very old
millisecond pulsars, in which non-thermal magnetospheric X-ray production
mechanisms are believed to dominate (see e.g. Becker \& Tr\"umper 1999, Becker
et al.~2003, Webb et al.~2004).
The old field pulsars aid in
answering questions such as "How do the emission properties of the younger
pulsars like Geminga, PSR 0656+14 and PSR 1055-52 change as they age from
$\sim 10^5$ to $10^7$ years"? Will the thermal emission simply fade away due to
cooling with increasing age or will the star be kept hot (at about $0.5-1
\times 10^5$ K) over millions of years due to energy dissipation by processes
such as internal frictional heating ($\dot{E}_{diss} \sim 10^{28}-10^{30}$
erg/s) and crust cracking, as proposed by Alpar et al.~(1984; 1998) and
Ruderman (1998) or by vortex creeping and pinning models (Shibazaki \&
Lamb 1989; Hirano and Shibazaki, 1997)? What happens to the non-thermal,
hard-tail emission seen in the X-ray spectra of the middle aged field
pulsars? Will this emission be the dominant source or will this component
also decay with time and will only thermal emission from the hot and
heated polar-caps remain?

Establishing whether thermal polar-cap emission is present, or not, is of
importance in confronting and comparing data with the many magnetospheric
emission models which predict hot polar caps (e.g.~Arons \& Scharlemann 1979;
Sturner \& Dermer 1995, Zhang \& Harding 2000; Harding \& Muslimov 2001; 2002;
2003). The heated polar cap is a consequence of pair-creation by the gap discharge,
after which a significant amount of highly energetic charged particles is
expected to stream back to the neutron star, heating the surface to a few
million degrees. Furthermore, many magnetospheric emission models
{\em require} hot-polar caps in order to lessen the work-function of
electrons and positrons in the surface down to $\sim 100$ eV and thus make it possible
to pull out enough electrons for the vacuum gap discharge/break-down
(cf.~Michel 1991). These models fail if there are no hot-spots. On the
other side, Sturner \& Dermer (1995) propose that high energy gamma-ray
photons are created in pulsar magnetospheres by inverse Compton scattering
of relativistic electrons and thermal photons. If this model is correct,
the absence of hot polar-caps (i.e.~the missing bath of thermal photons)
would imply that old -- but close-by and still powerful -- pulsars are not
$\gamma$-ray emitters, in agreement with the current observations.

There are further questions that involve the old pulsar's energy output.
The observed luminosity is but a small fraction of the total energy
available due to the rotation of the star. Where is the bulk of the
pulsar's spin-down power going to? Are there pulsar-wind nebulae around these
systems not previously detected due to the sensitivity limitations
imposed by previous satellites?

Up to now, only three old, non-recycled, pulsars have been detected and all
the detections were close to the sensitivity limits of the instruments, thus
strongly limiting the ability to explore the physical emission processes
at work in these neutron stars. The detected pulsars are PSR B1929+10
(Helfand 1983; Yancopoulos, Hamilton \& Helfand 1994), B0950+08 (Seward
\& Wang 1988; Manning \& Willmore 1994; Saito 1998) and B0823+26 (Sun et
al.~1993). All have a spin-down age of {\large $\tau$} $\sim 0.2-3 \times 10^7$
years; magnetic fields of about $B_\perp\sim 10^{11}-10^{12}$G; a close
distance of $d \sim 0.12 - 0.38$ kpc and a very small absorption column
of $N_H\sim 10^{18}-10^{20}\; \mbox{cm}^{-2}$ (see e.g.~Hobbs et al.~2003).
Spectral and temporal information was only available for the brightest
of these, PSR 1929+10. Its X-ray pulse profile is very broad with a
single pulse stretching across the entire phase cycle. The fraction of
pulsed photons is $\sim 30\%$ (Yancopoulos, Hamilton \& Helfand 1994). Its
X-ray spectrum, observed with ROSAT, could be equally well fit with a
power law (photon-index $\alpha\sim 2$) and a black-body spectrum
(thermal polar-cap emission, $T\sim 3.2\times 10^6$ K, $R_{bb}
\sim 20-50$\,m), leaving open the real nature of its emission (Becker \&
Tr\"umper 1997). Further evidence for a non-thermal nature of the pulsar's
X-ray emission was found by Saito (1998) based on ASCA data and more
recently by Wozna et al.~(2003) in a joint analysis of archival ROSAT
and ASCA data. Both, PSR 1929+10 and \PSRB\, are detected at optical 
wavebands (Pavlov et al.~1996; Zharikov et al.~2002; Mignani et al.~2003).

Making use of the large collection area, as well as the excellent timing,
spatial and energy resolution provided by XMM-Newton, the three pulsars \PSRB,
\PSRBB\, and \PSRJ\, which all belong to the sub-class of old rotation-driven
field pulsars, were observed as part of the guaranteed time and AO1 guest observer
program. We list the radio properties of these pulsars in Table~\ref{t:radio} and 
make the comments that follow.

Based on its spin-down age, \PSRB\ is the oldest field pulsar among the more 
than 50 rotation-driven pulsars detected in X-rays (see e.g.~Table 3 of 
Becker \& Aschenbach 2002 for a recent list of X-ray detected rotation-powered 
pulsars).  The pulsar's radio dispersion measure is among the smallest of all 
known radio pulsars and the dispersion-measure-based distance is is in good 
agreement with the $(262 \pm 5)$pc distance deduced from a parallax measurement 
(Brisken et al.~2002).  The low column density implies that it should be feasible 
to detect soft (below 0.5 keV) X-ray emission from \PSRB.

X-rays from \PSRB\, were first detected by Maning \& Willmore (1994) in a
$\sim 9\,\mbox{ksec}$ ROSAT PSPC observation ($\sim 55$ source
counts). These authors suggested that the emission either came from a
$T\sim (2.1\pm 0.2) \times 10^6$ K hot polar cap of $R_{bb}\sim 20$m or arose
from synchrotron or curvature radiation with a power law spectrum of photon-index
$\alpha=0.9^{+2.2}_{-1.3}$. Interpreting the soft X-ray emission entirely as
arising from the neutron star's surface, Becker (1994) computed a $3\sigma$
temperature upper limit of $T_s^\infty < 0.23 \times 10^6$ K assuming a 1.4
M$_\odot$ neutron star with a medium stiff equation of state.

\PSRBB\, has an inferred magnetic field strength
that is the highest among the X-ray-detected old field pulsars.
X-ray emission from \PSRBB\, was first discovered using ROSAT (Sun et
al.~1993). The ROSAT PSPC count rates were $0.0015 \pm 0.0004$ cts/s and
$0.0009 \pm 0.0003$ cts/s for the $(0.8-2.0$ keV) and $(0.5-2.0$ keV) bands,
respectively. The small number of detected source counts did not
allow one to identify the emission process but the data were in
agreement with the hypothesis that the X-rays were emitted from a
$\sim 1.8\times 10^6$ K hot thermal polar cap of size $R_{bb}\sim 140$m. If
the soft X-rays were assumed to arise entirely from the neutron star's surface,
a $3\sigma$ temperature upper limit is $T_s^\infty <0.34 \times 10^6$ K assuming 
a 1.4 M$_\odot$ neutron star with a medium stiff equation of state (Becker 1994).

\PSRJ\, (Thorsett et al.~1994; Camilo \& Nice 1995; Ray et al.~1996) is
another important representative of the group of old, but non-recycled,
pulsars. Its rotation period is the shortest among
the old rotation-powered field pulsars. Compared to \PSRB\ and \PSRBB , its
spin-down energy is two orders of magnitude higher whereas its inferred 
magnetic field strength is similar to that computed for \PSRB. The pulsar's 
spin-down age makes it intermediate between the older pulsars \PSRB\ \& \PSRBB\ 
and the cooling neutron stars, which all have a spin-down age of some hundred 
thousand years (see Table 3 of Becker \& Aschenbach 2002). Its radio 
dispersion-measure inferred distance suggest a medium to low neutral hydrogen 
column density. The moderate column density is an advantage in searching for 
X-ray emission from this pulsar, which has never been previously observed 
by any high energy mission and was not known to emit X-rays. \PSRJ\ is 
located about $1^\circ$ outside the boundary of the Cygnus-Loop supernova 
remnant but an association would require that the pulsar's true age is rather 
different compared with its spin-down age and that it was born with a very 
high velocity.

In \S2 we describe the XMM-Newton observations of \PSRB, \PSRBB\, and \PSRJ\,
and provide the details of the data processing and data filtering. The results
of the spectral and timing analysis are given in \S3 - \S5. We provide a summary
and concluding discussion in \S6.

\section{XMM-NEWTON OBSERVATIONS AND DATA REDUCTION\label{obs}}

PSR B0950+08 was observed with XMM-Newton on May $8-9$, 2002 (XMM revolution 442) for
a total on-source time of 101\,168 s. The MOS1 camera was operated in imaging
(PrimeFullWindow) and timing (FastUncompressed) mode for 53.2 ksec and 27.8 ksec,
respectively. The MOS2 camera was used in imaging mode exclusively for 83.7 ksec.
The EPIC-PN exposure time was 82.9 ksec. The EPIC-PN camera was set to operate
in small-window mode to provide imaging, spectral and timing information with a
temporal resolution of 5.67 ms. The higher temporal resolution of the EPIC-PN
is achieved at the cost of a 30\% higher detector dead time and a reduced field
of view of $4.4 \times 4.4$ arcmin (see e.g.~Becker \& Aschenbach 2002 for a
summary of XMM instrument modes suitable for pulsar studies). The medium filter
was used for the EPIC-PN and the MOS1/2 in all exposures of \PSRB.

\PSRBB\, was observed on April 26, 2002 (XMM revolution 436) for a total on-source
time of $\sim 50$ ksec. The instrument configuration was similar to that used
for observing \PSRB. The observation of \PSRJ\, was performed on November $21-22$,
2002 (XMM revolution 541) for a total on-source time of 17 ksec. For these observations
the MOS1/2 cameras were both operated in imaging mode and the EPIC-PN camera was setup
to work in small-window mode with the thin filter. We found that this setup guaranteed
a much more efficient use of the XMM-EPIC detectors as opposed to swapping the MOS1
from imaging to timing mode during the observations. A summary of exposure times and
instrument modes used for each of the observation is given in Table ~\ref{observations}.

XMM-Newton data have been seen to show timing discontinuities in the photon
arrival times with positive and negative jumps of the order of one to several
seconds (Becker \& Aschenbach 2002; Kirsch et al.~2003). Inspecting the log-files
from our processing of raw data we found that the EPIC-PN data of \PSRB\, exhibited
clock discontinuities showing three positive jumps of 1s randomly distributed over
the observation. We found no evidence for clock discontinuities in the \PSRBB\,
and \PSRJ\, data. We therefore used the SAS development track software
(xmmsas\_20040212\_2158-dt) which corresponds to the beta-release of XMM-SAS
Version 6.0 for the analysis of the EPIC-PN data and  which
detects and corrects most timing discontinuities during data processing. In
addition, known timing offsets due to ground station and space craft clock
propagation delays are corrected by this software. The MOS1/2 data of
\PSRB, \PSRBB\, and \PSRJ\, were analyzed using XMM-SAS Version 5.4.1.

Data screening for times of high background was done by inspecting the
lightcurves of the MOS1/2 and PN data at energies above 10 keV. Strong X-ray
emission from soft proton flares are seen in the observations of \PSRB\, and
\PSRBB. Creating the lightcurves with bins of 100s, we rejected those bins
where the MOS1/2 lightcurves had more than 140 cts/bin. For EPIC-PN data
we rejected times with more than 13 cts/bin. The observation of \PSRJ\, was
not affected by times of high background so that we could use all MOS1/2 data,
but we still rejected data with more than 13 cts/bin for the EPIC-PN data analysis.

For \PSRB\, the data screening reduced the effective exposure time for the
imaging modes of MOS1 and MOS2 to 19.8 ksec and 65.7 ksec, respectively. For
the EPIC-PN data the effective exposure time was reduced to 48.9 ksec. The
fast timing mode data of MOS1 in which, for the sake of a higher temporal
resolution, all events from the central MOS1-CCD are collapsed into a 1D-image,
were not considered as these data are superseded in quality by the EPIC-PN
small-window data. For \PSRBB\, and \PSRJ\,
the effective exposure times for the MOS1/2-CCDs are 65.6 ksec and 33 ksec,
respectively. The larger dead time, as well as the applied filter criteria,
reduced the effective exposure time for the EPIC-PN observations of \PSRB\,
and \PSRJ\, to 33.9 ksec and 10.8 ksec, respectively.

For the spectral analysis based on the MOS1/2 data we used only those events
with a detection {\em pattern} between $0-12$ (i.e. single, double and triple
events) and the {\em flag} parameter set to less than or equal to 1. The latter
criterion excludes events which are located near to a hot pixel, or to a bright
CCD column, or which are near to the edge of the CCD. For the EPIC-PN timing and
spectral analyzes, we used only single and double events,
i.e.~those which have a pattern parameter of less than, or equal to, 4 and a
flag parameter equal to zero. The energy range of the MOS1/2 and EPIC-PN
CCDs was restricted to $0.3-10$ keV. The accuracy of the detector
response matrix towards softer energies does not support the inclusion of
photons below 0.3 keV.

Because of the low counting rates, the RGS data were found to be of limited
use and was ignored for the purpose of this paper. \PSRB\, has an optical
counterpart at $\sim 26-27\,\mbox{mag}$ (Zharikov et al.~2002). This is,
however, $\sim 5-6$ orders of magnitudes fainter than the limiting sensitivity
of the optical monitor (OM) aboard XMM-Newton. Similar optical fluxes are
expected for \PSRBB\, and \PSRJ\, so that we also do not report on the
analysis and processing of the OM-data.


\section{\PSRB\label{PSRB}}

\begin{figure}
\centerline{\psfig{figure=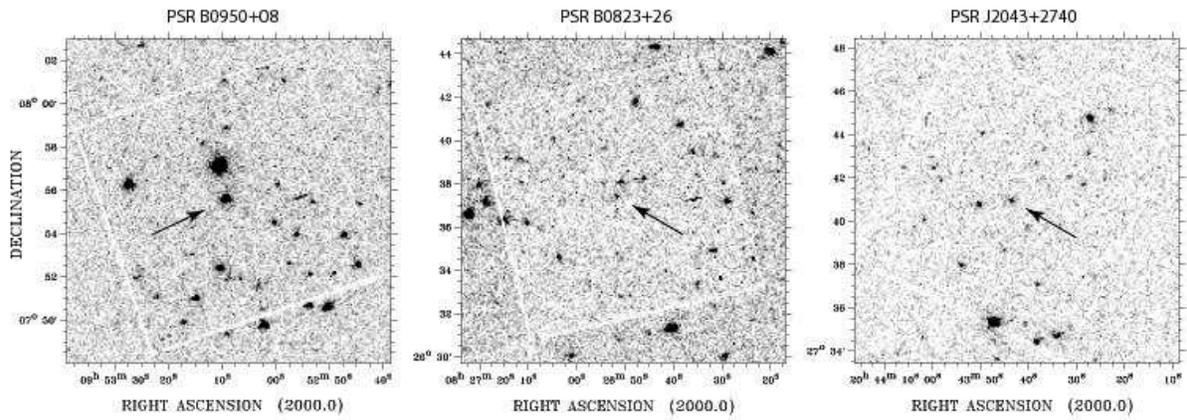,width=16cm,clip=}}
\caption[]{XMM's view of the $\sim 14 \times 14$ arcmin sky region around \PSRB,
\PSRBB\, and \PSRJ. Data from the MOS1 and MOS2 detectors have been merged to 
produce the images. The pulsars are indicated by an arrow. The bright source
near to \PSRB\, is  RX J095310.4+075712.}
\label{PSRB_MOS_field} \label{PSRBB_MOS_field} \label{PSRJ_MOS_field}
\end{figure}

The X-ray counterpart of \PSRB\, is detected with high significance in both
the MOS1/2 and EPIC-PN data. The counting rates are $0.0083 \pm 0.0009$ cts/s
(MOS1/2) and $0.029 \pm 0.001$ cts/s (EPIC-PN) within the $0.3-10$ keV band.
The maximum-likelihood source-detection did not yield any evidence for a
spatial extent of the pulsar's counterpart of larger than 15 arcsec,
corresponding to the HEW (Half Energy Width) of the instruments' point spread
function.  An $\sim 14' \times 14'$ image of the pulsar field based on the
merged MOS1/2 data is given in Figure \ref{PSRB_MOS_field}.

\subsection{Spectral Analysis\label{PSRB_spectral}}

The pulsar's energy spectrum was extracted from the MOS1/2 data by selecting
all events detected in a circle of radius 50 arcsec centered on the pulsar
position. Using the XMM-Newton/EPIC-MOS model point spread function, 90\% of
all the events from \PSRB\, are within this region. The presence of the
X-ray source RX J095310.4+075712 located $\sim 1.5$ arcmin near to the pulsar
(cf.~Figure \ref{PSRB_MOS_field}) precluded the extraction of the background
spectrum from an annulus surrounding the pulsar. We therefore extracted the
background spectrum from a source free region of 60 arcsec radius near to
the pulsar at RA(2000) $09^h\, 53^m\, 06.43^s$, DEC $07^\circ\, 53'\, 49.12\arcsec$.

For the EPIC-PN data we used an extraction radius of 33 arcsec radius, centered
on the pulsar. This selection region includes 85\% of the point source flux. As
the aim point is relatively close to the edge of the PN-CCD \#4, we extracted
the background spectrum from a source free region about one arc-minute east of
the pulsar at RA(2000) $09^h\, 53^m\, 12.75^s$, DEC $07^\circ\, 55'\, 06\arcsec$.
Although out-of-time events are small in the PN-timing mode, they prevent one
from extracting background spectra from a region located below the source and
along the CCD read-out direction.

In total, the extracted spectra include 1458 EPIC-PN source counts and 707
EPIC-MOS1/2 source counts. The spectral data were dynamically binned so as
to have at least 25 counts per bin. Model spectra were then simultaneously
fit to both the EPIC-PN and MOS1/2 data.

\begin{figure}
\centerline{\psfig{figure=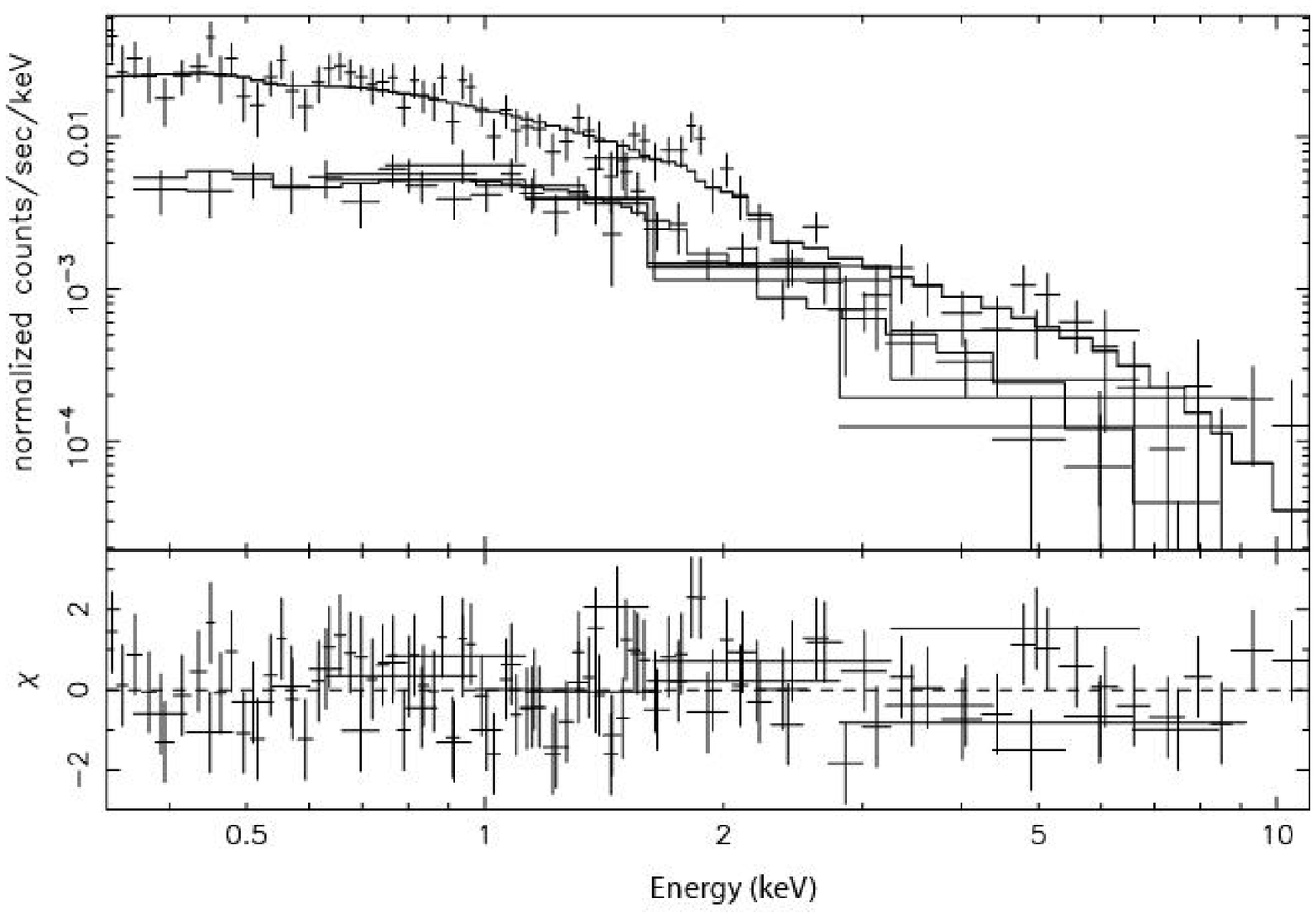,width=16cm,clip=}}
\caption[]{Energy spectrum of \PSRB\, as observed with the EPIC-PN (upper spectrum) 
and MOS1/2 detectors (lower spectra) and simultaneously fitted to an absorbed power 
law model ({\it upper panel}) and contribution to the \chisq\, fit statistic 
({\it lower panel}).} \label{PSRB_pl_spectrum}
\end{figure}

\begin{figure}
\centerline{\psfig{figure=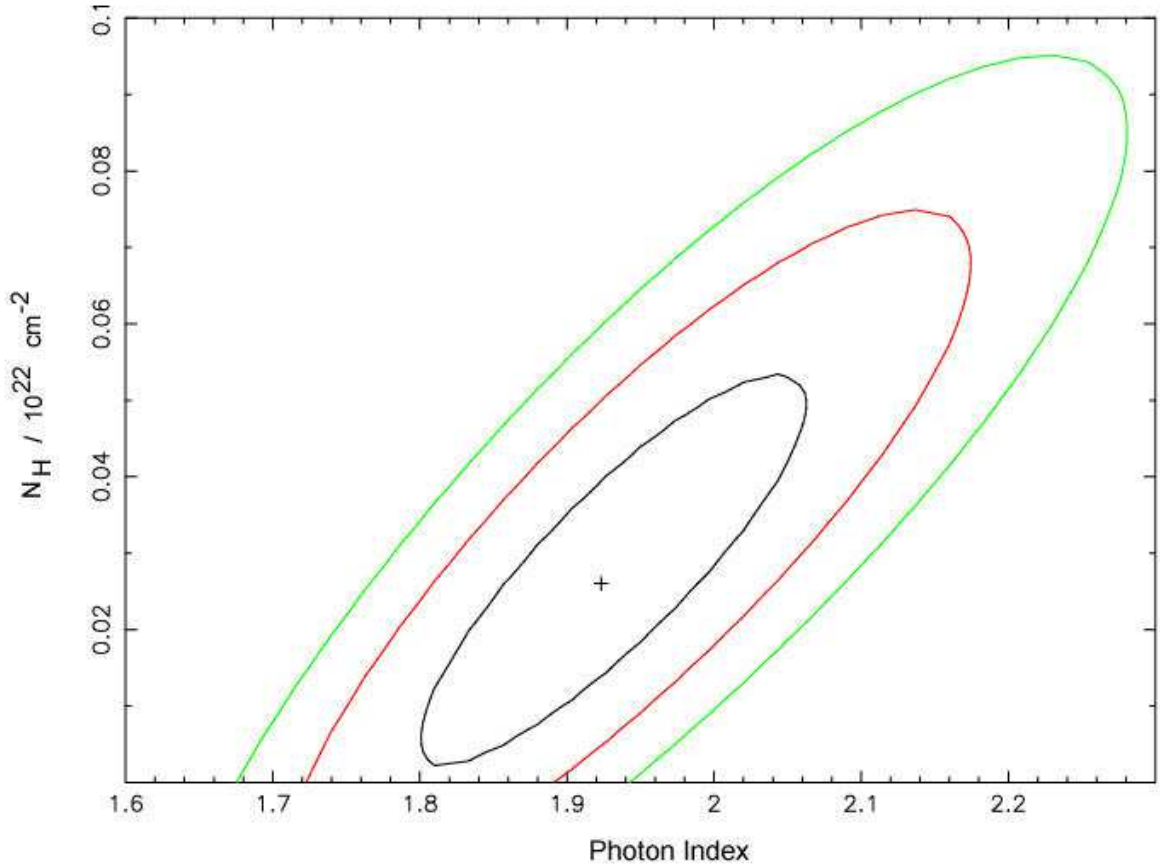,width=16cm,clip=}}
\caption[]{Contour plot showing the relative
parameter dependence of the photon index vs.~column absorption for the power law
fit to the \PSRB\, data. The three contours represent the $1-\sigma$, $2-\sigma$
and $3-\sigma$ confidence contours for two parameters of interest. The `+' sign
marks the best fit position, corresponding to $\chi^2_{min} =97.58$ for 112 dof.}
\label{PSRB_pl_contour}
\end{figure}

Amongst the single component spectral models, a power law model was found to
give the statistically best representation (\chisq =97.5 for 112 dof)
of the observed energy spectrum. A single blackbody (\chisq =280 for 112
dof) or a composite spectral model consisting of two blackbody components
(\chisq =279 for 110 dof) did not give acceptable fits. The best-fit power
law spectrum and residuals are shown in Figure~\ref{PSRB_pl_spectrum}. Contour
plots showing the relationship between the photon index and the column
absorption for various confidence levels are shown in Figure \ref{PSRB_pl_contour}.

The power law model yields a column absorption of $N_H=2.6^{+2.7}_{-2.4}\times
10^{20}\,\mbox{cm}^{-2}$, a photon-index $\alpha = 1.92^{+0.14}_{-0.12}$
and a normalization of $1.69^{+0.13}_{-0.11} \times10^{-5}$ photons cm$^{-2}$
s$^{-1}$ keV$^{-1}$ at $E=1$ keV. The errors represent the $1-\sigma$
confidence range computed for two parameters of interest. The column density
is in fair agreement with the neutral hydrogen column density $9.16 \times
10^{19}\,\mbox{cm}^{-2}$ deduced from the radio dispersion measure. For the
unabsorbed energy flux we measured
$f_x= 8.69^{+1.04}_{-0.90}\times 10^{-14}\, \,{\rm ergs\,\, s}^{-1}\,{\rm cm}^{-2}$
in the $0.5-10$ keV band, yielding an X-ray luminosity of $L_x=1.78_{-0.18}^{+0.22}
\times 10^{29}\,{\rm ergs\,\, s}^{-1}$. For the ROSAT energy band, $0.1-2.4$ keV,
we measured the flux to be $f_x=8.23^{+0.82}_{-0.59}\times 10^{-14}\,\,
{\rm ergs\,\, s}^{-1}\,{\rm cm}^{-2}$, yielding an X-ray luminosity of
$L_x= 1.69_{-0.12}^{+0.17} \times 10^{29}\, {\rm ergs\,\, s}^{-1}$.
These luminosities imply a rotational energy to X-ray energy conversion
factor of $L_x/\dot{E}= 3.18\times 10^{-4}$ within $0.5-10$ keV and
$2.85 \times 10^{-4}$ if transformed to the ROSAT band.

Energy spectra of cooling neutron stars like Geminga (Caraveo et al.~2004),
PSR 0656+14 (Greiveldinger et al 1996; Kennea, Becker \& Cordova 2004) and
PSR 1055-52 (Becker \& Aschenbach 2002) are found to be adequately represented
by spectral models consisting of three different components. Two blackbody
components representing the surface cooling and hot polar cap emission and
a non-thermal component dominating the energy range beyond $\sim 2$ keV.

Clearly, our analysis of the spectral data from \PSRB\, does not require all
three components. We therefore tried fitting the data with a composite model
consisting of a blackbody and a power law. The blackbody spectral component
then represents either thermal emission from a heated polar cap or residual
cooling emission from the whole neutron star surface whichever is more dominant.

Leaving free all fit parameters resulted in a model spectrum in which the
power law parameters were similar to those found in the single component
power law fit. The thermal component had a temperature of $\sim 6.3 \times 10^6$
K and a normalization of $R^2_{bb,km}/d^2_{10kpc} \sim 9.3\times 10^{-3}$. For the
pulsar located at a distance of $d=262$ pc, this corresponds to an emission radius
of $R_{bb}=2.5$ m which is unphysically small to be acceptable.

For comparison, defining the size of the presumed polar cap as the foot points of the
neutron star's dipolar magnetic field, the radius of the polar cap area is given
by $\rho=\sqrt{2\pi R^3/c P}$ with $R$ being the neutron star radius, c the
velocity of light and P the pulsar rotation period (see e.g.~Michel 1991).
For \PSRB\, with a rotation period of 253 ms this yields a polar cap radius
of $\rho\sim 287$ m, i.e.~more than hundred times the value obtained in the
spectral fit.

\begin{figure}

\centerline{\psfig{figure=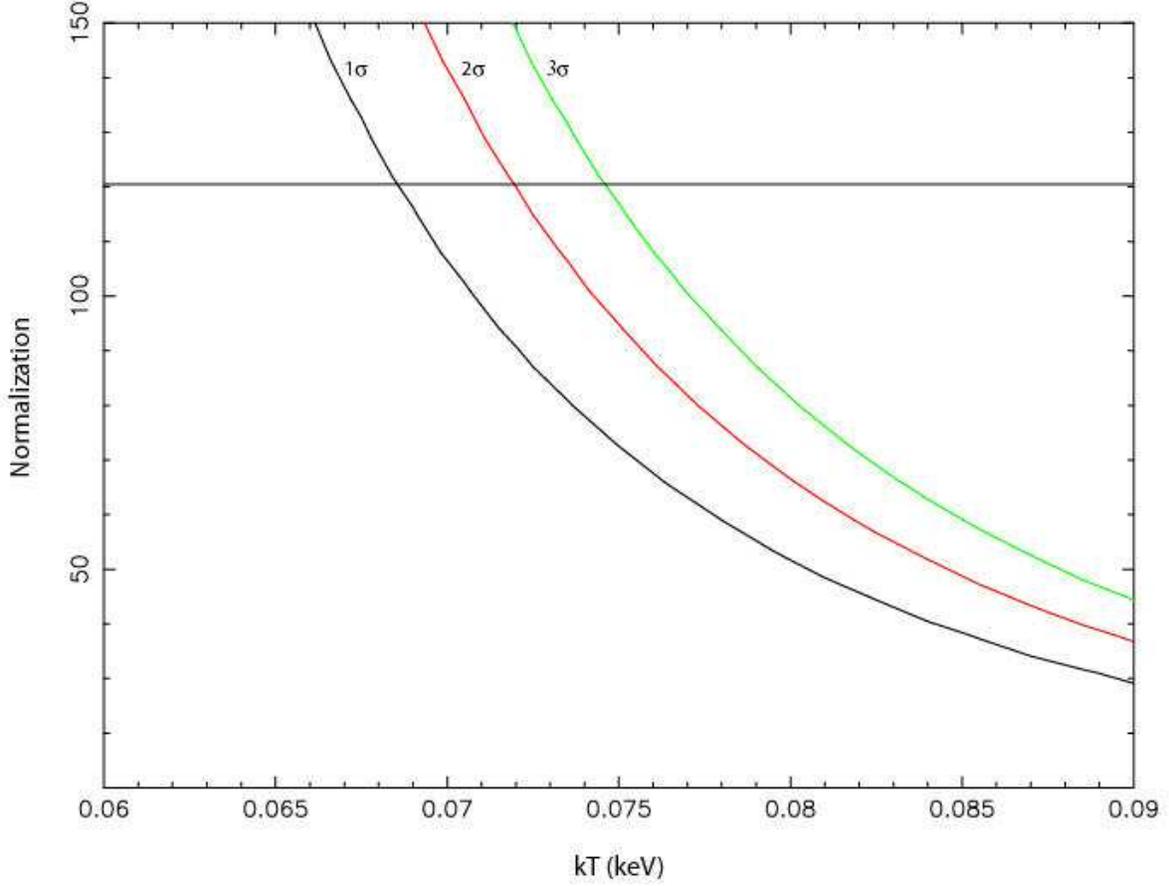,width=16cm,clip=}}
\caption[]{Portion of the confidence contours showing the blackbody normalization
versus blackbody temperature for the composite model (see text). The
horizontal line at a normalization of 120.6 corresponds to a polar cap radius
of $287$ m and a pulsar distance of 262 pc. The contours correspond to
$\chi^2_{min}=98.5$ plus 2.3, 6.17 and 11.8 which are the $1-\sigma$, $2-\sigma$
and $3\sigma$ confidence contours for 2 parameters of interest.}
\label{PSRB_bb_polarcap_pl_contour}
\end{figure}

As the thermal spectral component contributes mostly below $\sim 1$ keV, the
fitted column absorption is found to be a steep function of the blackbody
emitting area (normalization) and temperature. To determine
polar cap and surface temperature upper limits which are in agreement with
the observed energy spectrum and column absorption we fixed the absorption
of the composite model to the upper bound set by the $1-\sigma$ confidence
range deduced in the single power law fit. We then computed the confidence
ranges of the blackbody normalization and temperature by leaving all other
parameters free. The resulting contours, computed for two parameters of
interest, are shown in Figure \ref{PSRB_bb_polarcap_pl_contour}. For a
polar cap size of radius 287 m we can set a $3\sigma$ polar cap temperature
upper limit of $T_{pc}^\infty < 0.87 \times 10^6$ K assuming a contribution
from one polar cap only. For the thermal contribution coming from two polar
caps the $3\sigma$ temperature upper limit is $T_{pc}^\infty < 0.77 \times
10^6$ K. If we allow the thermal emission to be emitted from the whole
neutron star surface of 10 km radius we find a $3\sigma$ surface temperature
upper limit of $T_s^\infty < 0.48 \times 10^6$ K.

\subsubsection{Multi-wavelength Spectrum\label{PSRB_multi_spec}}

The results of a UBVRI photometry of \PSRB, using the FORS1 at the VLT UT1,
were recently reported by Zharikov et al.~(2003). The authors measured
the pulsar's optical magnitude in the different spectral bands to be 
B$=27.06 \pm 0.35$, V$=27.05\pm0.15$, $\mbox{R}_c=26.49\pm 0.10$ and 
$\mbox{I}_c=26.20\pm0.17$ (Johnson-Cousins system). A spectral fit to these 
data revealed a non-thermal spectrum with a photon-index of $1.65 \pm 0.4$,
albeit with large residuals (Zharikov et al.~2003). In order to investigate 
the pulsar's broadband spectrum we adopted the optical magnitudes from 
Zharikov et al.~(2003) and converted them to a monochromatic photon flux 
in the respective energy bands. 
Extrapolating the power law spectrum which describes the XMM-Newton data 
to the optical bands yields a photon flux which exceeds the measured one by 
more than an order of magnitude. This suggest that the broadband spectrum, 
if entirely non-thermal, has to break somewhere before or in the soft channels 
of the X-ray spectrum. We employ a broken power law model to parameterize 
the break between the optical and X-ray data. This model yields an acceptable 
description of both data sets (\chisq = 117 for 115 dof) with the break point 
fitted at $E_{break}= 0.67^{+0.18}_{-0.41}\,\mbox{keV}$.
The photon-index for $E < E_{break}$ and $E > E_{break}$ is found to be
$\alpha_1= 1.27^{+0.02}_{-0.01}$ and $\alpha_2 =1.88^{+0.14}_{-0.11}$,
respectively, with a normalization of $1.93 \times10^{-5}$ photons cm$^{-2}$ 
s$^{-1}$ keV$^{-1}$ at 1 keV. The column absorption is fitted to be 
$N_H=0.0^{+4.2}\times 10^{20}\,\mbox{cm}^{-2}$. The errors represent the 
$1-\sigma$ confidence range computed for two parameters of interest. 

\begin{figure}
\centerline{\psfig{figure=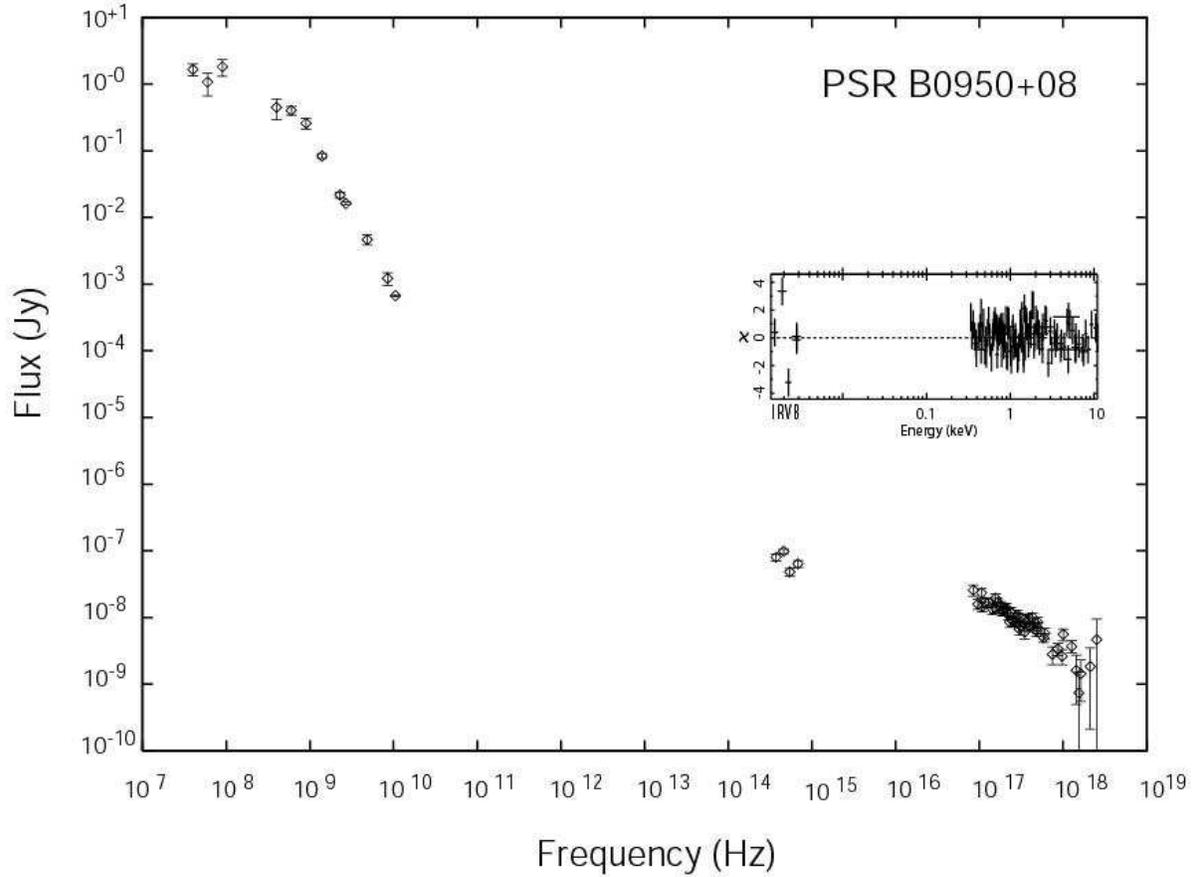,width=16cm,clip=}}
\caption[]{Combined radio, optical and X-ray spectral data of \PSRB.
The inset shows the contribution to the \chisq\, fit statistic for a
broken power law model fitted to the optical and X-ray data (see 
section \ref{PSRB_multi_spec} for further details).}
\label{PSRB_broadband_spectrum}
\end{figure}

In order to construct a broadband spectrum combining all spectral information available
from \PSRB\, we adopted the radio spectrum from Malofeev et al.~(1994) and plotted
it in Figure \ref{PSRB_broadband_spectrum} together with the optical and X-ray spectral 
data from the VLT and XMM-Newton. The inset depicts the errors from fitting the broken
power law model to the optical and X-ray data. The residuals for the fit to the optical
data from the R- and V-filters are of the order of $\sim 3 \sigma$ as found from the 
single power law fit by Zharikov et al.~(2003). The radio spectrum is not steep 
($\alpha= 1.2 \pm 0.3$) up to 2~GHz but then steepens to $\alpha=-2.4\pm 0.3$. 
The flux in the radio part of the spectrum, which is supposed to be due to coherent 
radiation, is several orders of magnitude greater than the extrapolated optical 
or X-ray fluxes.

\subsection{Timing Analysis\label{PSRB_timing}}

We used the EPIC-PN small-window mode data for the timing analysis.
The temporal resolution of 5.67ms, available in this mode, is
more than sufficient to resolve the 253 ms period from \PSRB.
Events were selected from a circle of 25 arcsec radius centered on
the pulsar. This extraction region contains 80\% of the point source
flux. For the barycenter correction we applied the
standard procedures for XMM-Newton data using {\em barycen-1.17}\/
and the JPL DE200 Earth ephemeris to convert photon arrival times
from the spacecraft to the solar system barycenter (SSB) and the
barycentric dynamical time (TDB). The pulsar's radio position,
RA(2000) $09^h\, 53^m\, 9.307^s$, DEC $07^\circ\, 55'\, 35.93\arcsec$,
was used for the barycenter correction.

\PSRB\, is regularly monitored in the radio band using the Max-Planck 
100m radio telescope in Effelsberg. In addition, pulsar ephemerides from 
the Princeton radio pulsar database\footnote{The Princeton radio pulsar 
database is available from ftp://pulsar.princeton.edu/gro/psrbin.dat} 
were used for the analysis. The pulsar's spin-parameters $P$ and $\dot{P}$ 
are therefore known with high precision and can be extrapolated with 
sufficient accuracy to the mean epoch of the XMM-Newton 
observation: MJD=52403.2576249456 (TDB@SSB). \PSRB\, is not known to 
show timing irregularities (glitches) so that we can fold the photon 
arrival times using the extrapolated rotation frequency and frequency 
time derivative.

The statistical significance of the pulsations was computed using the
$Z^2_n$-test with $1-10$ harmonics in combination with the H-Test
to determine the optimal number of harmonics (De Jager 1987; Buccheri
\& De Jager 1989). The optimal number of phase
bins for the representation of the pulse profile was computed by taking
into account the signal's Fourier power and the optimal number of
harmonics deduced from the H-Test (see Becker \& Tr\"umper 1999).

Within the $0.3-10$ keV energy band, 1730 events were available for
the timing analysis of which $\sim 20\%$ is estimated to be background.
The $Z^2_n$-test gave 52.96 with 3 harmonics. According to the H-Test,
the probability of measuring this quantity by chance is $1.2 \times
10^{-9}$ thus establishing \PSRB\, firmly as an X-ray pulsar.

By restricting the timing analysis to the $0.3-2.0$ keV and $2.0-10$ keV
energy bands, respectively, we found that most of the power of the pulsed
emission is coming at energies below 2.0 keV. The number of counts available
in this bands are 1328 and 378, respectively, with a 12\% and 40\% background
contribution. Within the $0.3-2.0$ keV band the $Z^2_n$-test gave 50.83 for
n=3 harmonics (corresponding  to $3.2 \times 10^{-9}$ chance occurrence)
whereas in the $2-10$ keV band we found $Z^2_n=10.15$ for n=2 harmonics
(corresponding to $3.8 \times 10^{-2}$ chance occurrence).

We computed the fraction of pulsed events by using a bootstrap method
(Swanepoel, de Beer \& Loots 1996; Becker \& Tr\"umper 1999). For the
energy range  $0.2-10$ keV we find $28\pm 6\%$.  The fraction of pulsed
events in the $0.3-2.0$ keV and $2.0-10$ keV energy bands are $30 \pm 6\%$
and $24 \pm 12\%$, respectively.

\begin{figure}
\centerline{\psfig{figure=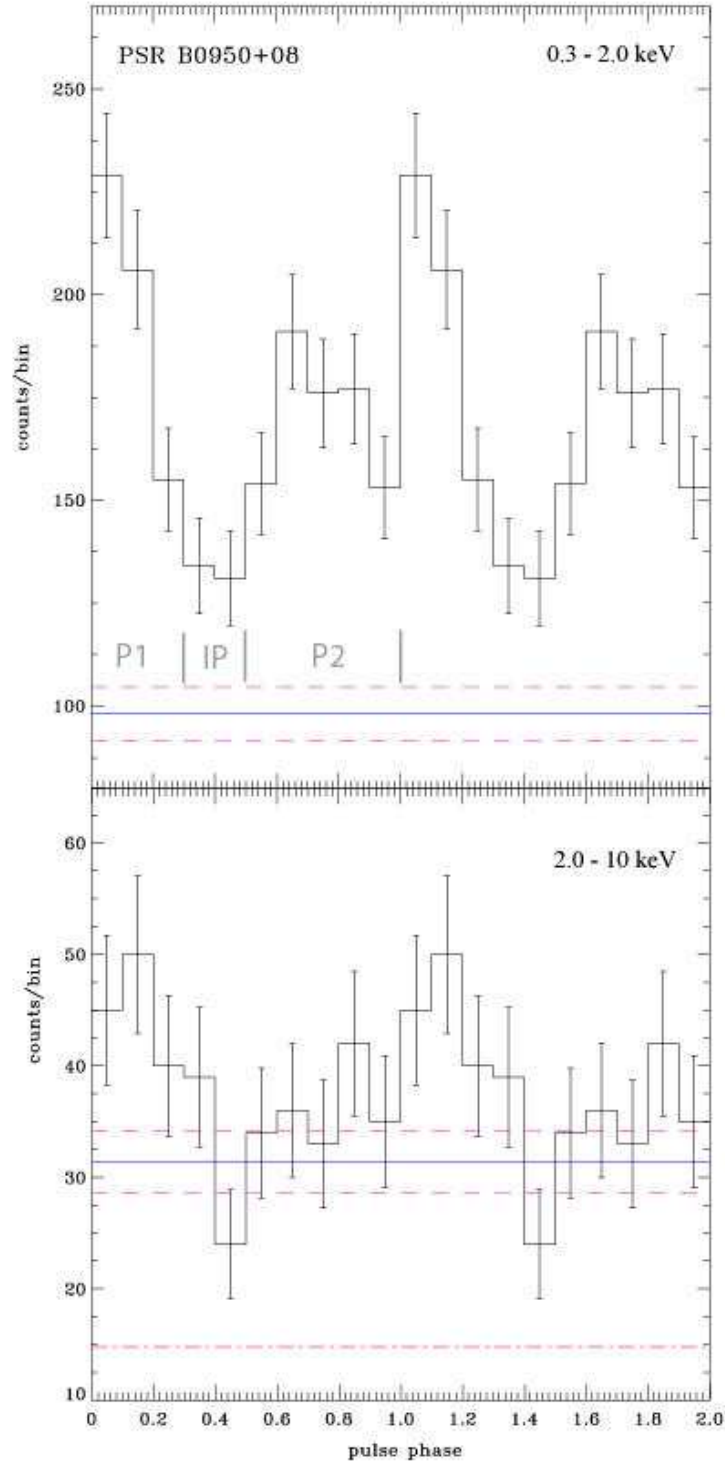,height=20cm,clip=}}
\caption[]{Integrated pulse profiles of \PSRB\, as observed with the XMM-Newton
EPIC-PN in the $0.3-2.0$ keV and $2.0 - 10$ keV energy bands. Two phase cycles
are shown for clarity. The solid and dashed lines indicate the DC level and
its uncertainty range. The dashed-dotted line represents the background contribution.
The phase ranges for peak one (P1), peak two (P2) and the interpulse (IP) are indicated.}
\label{PSRB_pulseprofiles}
\end{figure}

\begin{figure}
\centerline{\psfig{figure=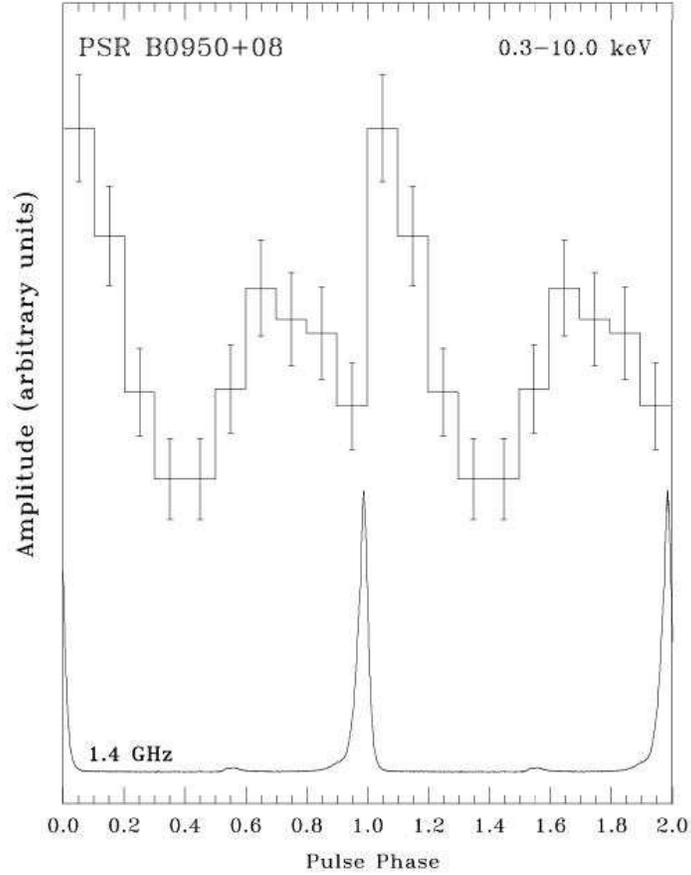,height=12cm,clip=}}
\caption[]{Integrated pulse profiles of \PSRB\, as observed in the $0.3-10$ keV
band (top) and and at 1.4 GHz with the Effelsberg radio telescope (bottom). 
X-ray and radio profiles are phase related. Phase zero corresponds to the mean
epoch of the XMM-Newton observation. Two phase cycles are shown for clarity.} 
\label{PSRB_x_radio_profiles}
\end{figure}

Figure \ref{PSRB_pulseprofiles} depicts the X-ray pulse profile of \PSRB\,
for the $0.3-2.0$ keV and $2.0-10$ keV energy bands.  As indicated by the
higher harmonic content and easily seen in the figure, the pulse shape is
not sinusoidal but shows two pulse peaks with a separation of $\sim 144^\circ$
(maximum to maximum) between the peaks. The X-ray pulse profile compared
with the radio lightcurve, taken at 1.4 GHz with the Effelsberg radio telescope,
is shown in Figure \ref{PSRB_x_radio_profiles}. The radio pulse profile shows
a similar doubly peaked structure as observed in X-rays, although the interpulse
component is much less intense than the main pulse. The phase separation
between the two radio peaks is $\sim 160^\circ$. The similarity of the phase 
and spacing of the radio and X-ray pulses suggests that the pulsed emission 
at radio and X-ray wavelengths originates from the same location in the magnetosphere.
Computing the absolute phase of the main radio pulse peak using observations
with Effelsberg made between $2001-2004$ revealed that the main radio peak leads 
the trailing X-ray peak by $\sim 0.06 - 0.14$ in phase, depending on if we 
take the maximum  or the center of mass of the peaks as fiducial point
(see Figure \ref{PSRB_x_radio_profiles}). The phase of the main radio peak 
at the epoch of the XMM observation is $0.98 \pm 4\%$
 
We note that in order to check the accuracy of the XMM-Newton clock against 
UTC we analyzed archival Crab-pulsar data taken in March 2002 and October 2003. 
The phase difference between the arrival of the Crab pulsar's main radio pulse 
and the X-ray pulse was fitted to be $\sim 300\mu $s and $\sim 275\mu $s in 
the March 2002 and October 2003 data, respectively. Assuming that the Crab 
pulsar X-ray pulse is phase aligned with the radio pulse the XMM-Newton clock 
accuracy is better than 1/80 phase bin in the X-ray pulse profile of \PSRB\, 
shown in Figure~\ref{PSRB_pulseprofiles}.

\subsection{Phase resolved spectral analysis}

In order to investigate some possible spectral variations as a function of
pulse phase $\phi$ we have performed a phase resolved spectral analysis.

Two approaches are common in this analysis, differing only by the definition 
of what events are taken for the background subtraction with the 
sky-and-instrument background selected from a source free region 
near to the pulsar and events from an apparent off-pulse region in the 
pulse profile as the usual choices. The later approach, however, suffers 
from a conceptional problem in that it needs a clear proof that the 
off-pulse events are indeed not part of the pulsed emission but caused
by a different radiation mechanism. This information, however, is usually
not available so that defining the interpulse region as off-pulse emission 
is not physically motivated. The Crab pulsar might be a good example to make 
this clear. Its off-pulse emission was usually taken as background emission 
in phase resolved spectral analysis but only recently it was found with {\em Chandra} 
that the pulsar is ''on'' throughout its pulse phase (Tennant et al. 2001; 
Weisskopf et al.~2004), invalidating the apparent off-pulse region as being 
DC background emission. Simulating pulse profiles makes clear that any apparent 
off-pulse region as well as DC level and pulsed fraction depend strongly on 
the viewing geometry under which a pulsar is observed (cf.~\S\ref{discussion} and 
Figure \ref{tpc}) and that these parameters are not inherent parameters 
which might be used to unambiguously characterize the emission process.

We have selected and labeled pulsar events 
according to Figure \ref{PSRB_pulseprofiles}, defining the phase region 
$0.0 \le \phi < 0.3$ as peak one (P1), $0.3 \le \phi < 0.5$ as the 
interpulse region (IP) and $0.5 \le \phi < 1.0$ as peak two (P2) and
corrected for the sky and instrument background using the same data as were
used for the phase averaged spectral analysis discussed in \S~\ref{PSRB_spectral}. 
Before subtracting background the number of events were 641 counts in P1, 961 counts 
in P2, and 277 counts in the interpulse phase range. All spectral data were dynamically 
binned so as to have at least 25 counts per spectral bin. The spectra from P1, P2 and 
IP were then fit to model spectra.

The phase resolved spectra for P1, P2 and IP support the non-thermal character
of the pulsar emission. Single blackbody spectra gave large residuals beyond 
$\sim 3-5$ keV with unphysically small normalizations (emitting areas).  In 
contrast, fitting a power law spectrum yields acceptable descriptions of the 
emission ($\chi^2_{P1}=13$ for 22 dof; $\chi^2_{P2}= 20 $ for 34 dof; $\chi^2_{IP}= 5$  
for 8 dof) in the full $0.3-10$ keV energy range with no significant spectral 
variations as function of pulse phase. For the power law index of P1, P2 and IP
we found $\alpha_{P1}=1.77^{+0.29}_{-0.27}$, $\alpha_{P2}=1.98^{+0.25}_{-0.23}$ 
and $\alpha_{IP}= 1.75^{+0.35}_{-0.27}$, respectively. The corresponding $1\sigma$
confidence ranges of the column density are $0.0-9.1 \times 10^{20}\, \mbox{cm}^{-2}$, 
$0.0-4.2 \times 10^{20}\, \mbox{cm}^{-2}$ and $0.0-5.2 \times 10^{20}\, 
\mbox{cm}^{-2}$, respectively, which makes the phase-resolved spectra fully 
in agreement with the results found in the analysis of the phase averaged spectrum. 
Thus, within the statistical uncertainties, there is no evidence for spectral variation 
as function of pulse phase.


\section{\PSRBB\label{PSRBB}}

Although \PSRBB\, was observed by XMM-Newton with an integration time only
about a factor of two less than \PSRB, the number of detected photons was
down by more than a factor of eight. We
measured an EPIC-PN source count rate of $0.0035 \pm 0.0003$ counts/s
($0.3-10$ keV). For the MOS1/2 detectors, the average rate
was $0.0011 \pm 0.0001$ counts/s. The source extent of
$\le 15$ arcsec (HEW) is in agreement with the point spread function
for an on-axis point source.  An $\sim 14' \times 14'$ image of the 
pulsar field based on the merged MOS1/2 data is given in 
Figure \ref{PSRB_MOS_field}.

\subsection{Spectral Analysis\label{PSRBB_spectral}}

For \PSRBB\ we used a similar approach as described for \PSRB. To
extract the pulsar spectrum from the EPIC-PN and MOS1/2 data,
we selected all events within a circle of 33 arcsec centered on the pulsar. For
the MOS1/2 data, the background spectrum was extracted from an annulus of 60
arcsec radius at RA(2000) $08^h\, 26^m\, 55.055^s$, DEC $+26^\circ\, 36'\, 42.17\arcsec$.
For the EPIC-PN data, we selected the background spectrum from a source free region
one arc-minute east of the pulsar at RA(2000) $08^h\, 26^m\, 47.178^s$, DEC $+26^\circ\,
37'\, 16.54\arcsec$. In total, the extracted spectra included 121 source counts from
the EPIC-PN and 70 source counts from the EPIC-MOS1/2. The data were binned to
achieve at least 20 counts per bin.

\begin{figure}
\centerline{\psfig{figure=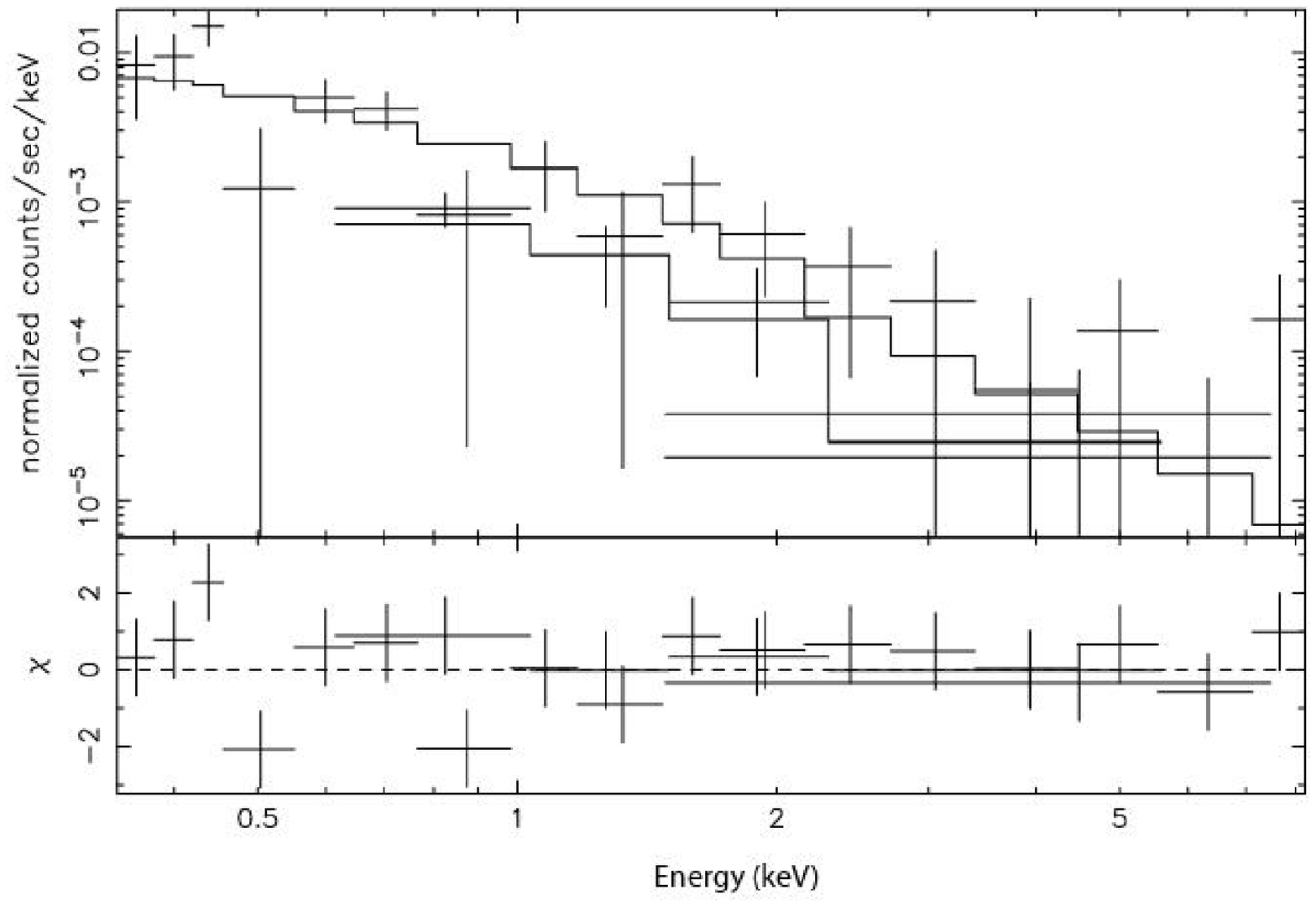,width=16cm,clip=}}
\caption[]{Energy spectrum of \PSRBB\, as observed with the EPIC-MOS1/2
and PN detectors and simultaneously fitted to an absorbed power law model
({\it upper panel}) and contribution to the \chisq\, fit statistic ({\it lower panel}).}
\label{PSRBB_spectrum}
\end{figure}

Among the various spectral models which were tested, we found that a single
blackbody model provided the worst description of the data ($\chi^2 =30.4$ for
19 dof). Including a second blackbody did not provide a
significant improvement ($\chi^2 =29.8$ for 17 dof). A single power law with
photon-index $\alpha=2.5^{+0.9}_{-0.45}$ was found to give the best description
($\chi^2 =20.6$ for 19 dof). The power law model yields a column absorption of
$N_H=0^{+8.8}\times 10^{20}\,\mbox{cm}^{-2}$ and a normalization of
$2.08^{+0.71}_{-0.53} \times10^{-6}$ photons cm$^{-2}$ s$^{-1}$ keV$^{-1}$
at $E=1$ keV. The errors are the $1-\sigma$ confidence range computed
for two parameters of interest. The best-fit power law spectrum and residuals
are shown in Figure \ref{PSRBB_spectrum}.

In the $0.5-10$ keV band, the power law model yields an unabsorbed energy flux of
$f_x= 7.33^{+0.96}_{-0.16}\times 10^{-15}\, \,{\rm ergs\,\, s}^{-1}\,{\rm cm}^{-2}$.
For a pulsar distance of 340 pc, this corresponds to an X-ray luminosity of
$L_x=1.01_{-0.02}^{+0.13} \times 10^{29}\,{\rm ergs\,\, s}^{-1}$. For
the ROSAT energy band we measured the flux to be  $f_x=2.68^{+6.28}_{-0.89}
\times 10^{-14}\,\,{\rm ergs\,\, s}^{-1}\,{\rm cm}^{-2}$, yielding an X-ray luminosity of
$L_x= 2.32_{-1.24}^{+6.28} \times 10^{29}\, {\rm ergs\,\, s}^{-1}$. These luminosities
imply a rotational energy to X-ray energy conversion factor of $L_x/\dot{E}= 2.21\times
10^{-4}$ and $5.08 \times 10^{-4}$ if transformed to the $0.1-2.4$ keV ROSAT band.

To test the possibility of some thermal contribution from the whole neutron star
surface, or part of it, we again fitted the data with a composite model consisting
of a thermal blackbody
and a power law. As for \PSRB, leaving free all fit parameters resulted
in a good fitting spectral model ($\chi^2 =20$ for 17 dof) in which the photon-index
and power law normalization were found to be similar to the values obtained in the
single component fit but with a blackbody emitting area of about one centimeter radius.
For comparison, using the same formula as given in \S\ref{PSRB_spectral} the polar
cap size of \PSRBB\, is computed to be $\rho=199$ m. The small blackbody normalization
obtained from the fits is a direct consequence of the fact that the power law model
already provides an acceptable description of the observed energy spectrum.

We therefore computed $3\sigma$ upper limits for the polar cap and surface
temperature using the same approach as for \PSRB. For a polar cap of radius
$\rho=199$ m we find a temperature upper limit of $T_{pc}^\infty < 1.17\times 10^6$ K.
If we allow the thermal emission to come from two polar caps of this size we
compute $T_{pc}^\infty < 1.08\times 10^6$ K for each of them. Assuming a
thermal contribution, not from a heated polar cap, but from the entire neutron
star surface of 10 km radius yields $T_s < 0.5 \times 10^6$ K.

\begin{figure}
\centerline{\psfig{figure=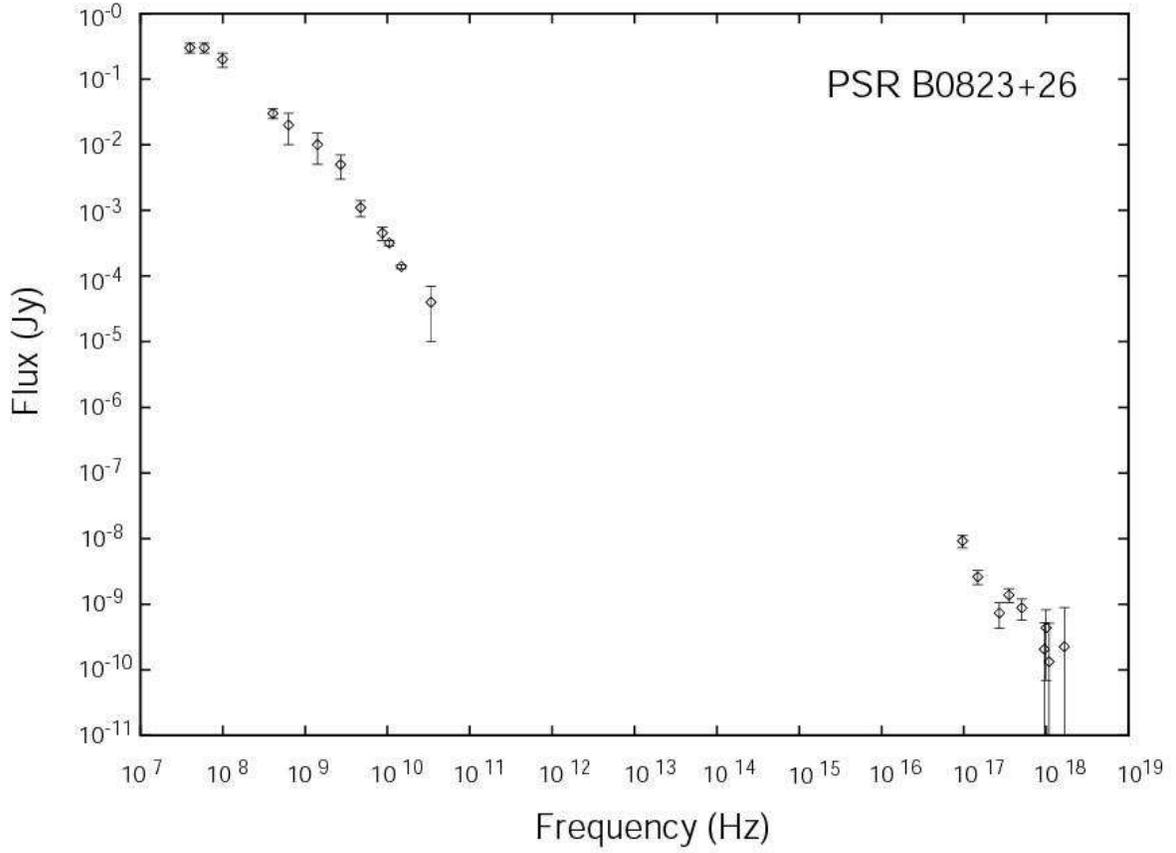,width=16cm,clip=}}
\caption[]{Combined radio and X-ray spectral data of \PSRBB.}
\label{PSRBB_broadband_spectrum}
\end{figure}

\PSRBB\ belongs to the small minority of known pulsars that are visible at high 
radio frequencies ($\nu > 30\rm GHz$). A radio spectrum was measured by  
Malofeev et al.~(1994). In order to construct a wideband spectrum using all spectral 
information available from \PSRBB\, we  converted the X-ray photon counts  to flux units
and combined  them in  Figure \ref{PSRBB_broadband_spectrum} with the radio fluxes.
The radio spectrum has a spectral index of  $\alpha = 1.3 \pm 0.1$ up to the break frequency 
of 4~GHz where the spectrum steepens slightly to $\alpha = 1.8 \pm 0.1$.  The origin of 
this high frequency emission is still unknown, but seems  to be more closely linked to 
the driving process of radio emission as all high frequency emitters show an abnormally 
strong, most likely intrinsic, intensity modulation at high frequencies.  In the case of  
\PSRBB\, the slopes of radio and X-ray spectra are similar, but the flux in the X-ray part 
amounts  more than $\sim 10^6$ times of the extrapolated radio spectrum.

\subsection{Timing Analysis\label{PSRBB_timing}}

\begin{figure}
\centerline{\psfig{figure=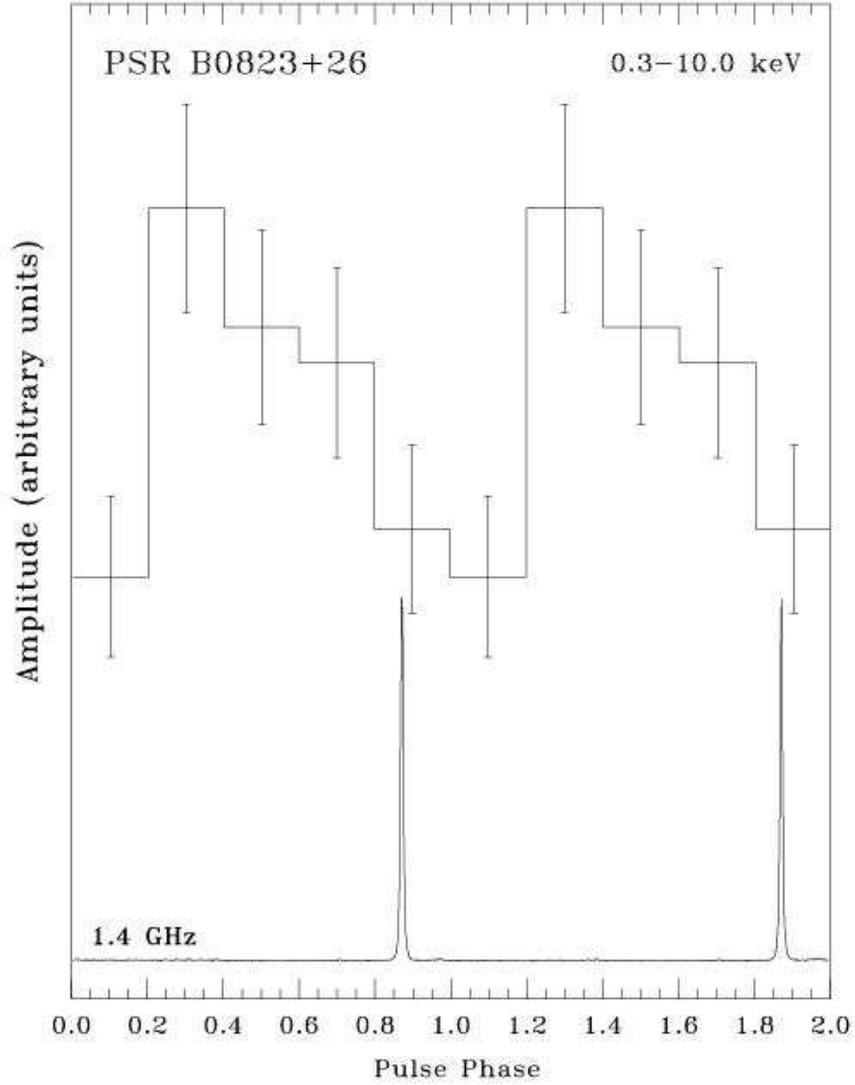,width=12cm,clip=}}
\caption[]{Integrated pulse profiles of \PSRBB\, as observed with the EPIC-PN
aboard XMM-Newton (top) and at 1.4 GHz with the Effelsberg radio telescope
(bottom). X-ray and radio profiles are phase related. Phase zero corresponds 
to the mean epoch of the XMM-Newton observation. Two phase cycles are 
shown for clarity.}
\label{PSRBB_pulseprofiles}
\end{figure}

For the timing analysis we selected all events detected in the EPIC-PN
within a circle of 25 arcsec centered on the pulsar. This yielded 304
events of which $\sim 64\%$ are background. \PSRBB\, is not observed to
show timing irregularities (glitches) and this allows us to simply
epoch-fold the photon arrival times using  pulsar ephemeris from the 
Princeton Pulsar Catalog as well as from Effelsberg radio data of \PSRB\, 
which we extrapolated to the mean epoch MJD=52391.0168077238 (TDB@SSB) of the 
XMM-Newton observation. The H-test indicated the highest probability 
$Z_n^2 = 6.97$ for $1$ harmonic. The test statistic $Z^2_n$ is distributed, 
in the absence of a signal, as $\chi^2$ with $2n$ degrees of
freedom. Thus the test statistic implies a probability of chance occurrence
of $\sim 3\%$, i.e. $\sim 2.2 \sigma$ {\em evidence} that the observed
signal is real, and not simply due to statistical fluctuations. The
corresponding pulse profile is shown in Figure \ref{PSRBB_pulseprofiles}.
The pulsed fraction is $49 \pm 22\%$ with the large error reflecting the 
low significance of the observed modulation. 

To compute the absolute phase of the main radio pulse at the epoch of
the XMM observation we used data taken with Effelsberg between $2001-2004$. 
From this data we found the phase of the radio peak at $0.87 \pm 4\%$
(see Figure \ref{PSRBB_pulseprofiles}).


\section{\PSRJ\label{PSRJ}}

The 96 ms pulsar \PSRJ\, was detected in the $0.3-10$ keV energy band
at a $\sim 10\sigma$ level of statistical significance, establishing
it as X-ray bright and making it one of the rare old, but non-recycled,
rotation-powered pulsars detected in X-rays. An image is
given in Figure \ref{PSRJ_MOS_field}. The source extent of $\le 15$
arcsec (HEW) is consistent with that expected for an on-axis point source.
The EPIC-PN source count rate is
$0.0085 \pm 0.0009$ counts/s in the $0.3-10$ keV band. The MOS1/2
count rate is $0.0012 \pm 0.0003$ counts/s.

\subsection{Spectral Analysis\label{PSRJ_spectral}}

\begin{figure}
\centerline{\psfig{figure=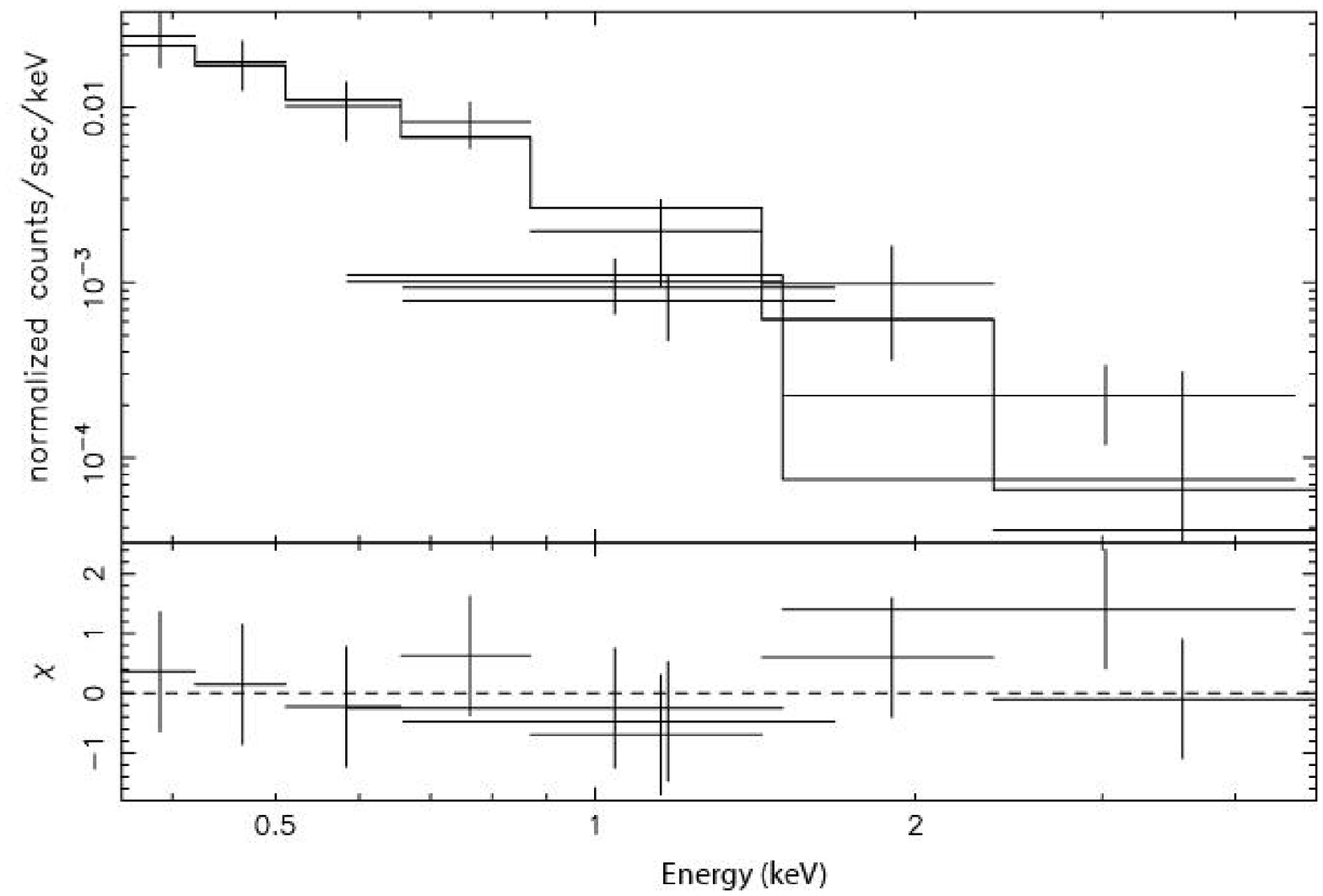,width=16cm,clip=}}
\caption[]{Energy spectrum of \PSRJ\, as observed with the EPIC-MOS1/2
and PN detectors and simultaneously fitted to an absorbed power law model
({\it upper panel}) and contribution to the \chisq\, fit statistic
({\it lower panel}).}
\label{PSRJ_spectrum}
\end{figure}

For \PSRJ\, we applied the same approach as described for \PSRB\, and
\PSRBB\, in sections \ref{PSRB} and \ref{PSRBB}. From the MOS1/2 data
the background spectrum was selected from an annulus of 34 and 70 arcsec
inner and outer radius, respectively, centered on the pulsar position.
In the EPIC-PN we extracted the background spectrum from a source free
region of 33 arcsec radius, centered at RA(2000) $20^h\, 43^m\, 47.44^s$,
DEC $+27^\circ\, 41'\, 17\arcsec$. The extracted spectra include 92 source
counts from the EPIC-PN and 40 source counts from the EPIC-MOS1/2. The
data were binned so that there were at least 20 counts per bin. Amongst
the tested models, a power law with photon-index $\alpha=3.1_{-0.6}^{+1.1}$
yields a better description of the observed energy spectrum than a
single blackbody  model does, albeit the errors of the fitted parameters
are large in both cases as shown in Table~\ref{spectral_fits_J2043}.
The observed energy spectrum fitted with an absorbed power law
model is shown in Figure \ref{PSRJ_spectrum}.

The quality of the power law model is sufficient to describe the observed
energy spectrum. To deduce upper limits for a thermal contribution we
again fit a composite model consisting of a power law and blackbody. The
polar cap size inferred for a 96ms pulsar is $467$ m. Using the same
fitting techniques as for the other two pulsars we computed a $3\sigma$
upper limit for the polar cap temperature of $T^\infty_{pc} < 1.45 \times
10^6$ K, and of $T^\infty_{pc} < 1.23 \times 10^6$ K if the thermal
emission is coming from two polar caps of the same size which are
contributing at the same time. If the thermal contribution is coming
from the whole neutron star surface of 10 km radius we found a $3\sigma$
surface temperature upper limit of $T^\infty_{s} < 0.627 \times 10^6$ K.
A summary of the spectral fitting is given in Table \ref{spectral_fits_J2043}.

As with \PSRB\, and \PSRBB\, we estimate the unabsorbed energy flux for the
pure non-thermal model. In the $0.5-10$ keV band this yields
$f_x\sim 1.1\times 10^{-14}\,\,{\rm ergs\, \, s}^{-1}\,{\rm cm}^{-2}$ and
for the ROSAT band we compute $f_x \sim7.2\times 10^{-14}\,\,{\rm ergs\,\, s}^{-1}
\,{\rm cm}^{-2}$. Assuming a distance of 1.8 kpc, the corresponding
luminosities are $L_x= 4.26 \times 10^{30}\, {\rm ergs\,\, s}^{-1}$ and $L_x= 2.8
\times 10^{31}\, {\rm ergs\,\, s}^{-1}$, respectively. These luminosities
imply a rotational to X-ray energy conversion factor of $L_x/\dot{E}= 7.6\times
10^{-5}$ for the $0.5-10$ keV energy range, or $L_x/\dot{E}=
5 \times 10^{-4}$ if converted to the ROSAT band.

\subsection{Timing Analysis\label{PSRJ_timing}}

196 events from the EPIC-PN, of which we estimate $\sim 43\%$ are from the
background, were used for timing analysis. We folded the barycenter-corrected
photon arrival times to the pulsar ephemeris which we extrapolated from the ephemeris
listed in the ATNF Pulsar Catalogue (Hobbs et al.~2003) to the mean epoch of the
XMM-Newton observation: MJD=52600.0643265238 (TDB@SSB). The H-test indicated the
highest probability $Z_n^2 = 3.3$ for 1 harmonic. The test statistic thus yields a
probability of chance occurrence of $\sim 19\%$. The observed modulation is
therefore not considered as evidence for pulsations. We computed a $2\sigma$ pulsed
fraction upper limit of $ 57 \%$ assuming a sinusoidal pulse profile.

\section{DISCUSSION \& SUMMARY\label{discussion}}

We have investigated the X-ray emission properties of three old, but non-recycled,
rotation-powered pulsars in order to probe and identify the origin of their
X-radiation.  These pulsars, being intermediate in age between the young
cooling neutron stars and the old recycled millisecond pulsars, are of special interest
as they provide important information for understanding the X-ray emission properties
of rotation-powered pulsars as a class. The selected targets provide a valuable
snapshot at ages {\Large $\tau$}$ = 1.2 \times 10^6$ years (\PSRJ),
{\Large $\tau$}$ = 4.89 \times 10^6$ yrs (\PSRBB) and {\Large $\tau$}$ = 1.74
\times 10^7$ yrs (\PSRB) and allows one to add to the current picture of pulsar
X-ray emission properties beyond  the younger (spin-down ages between $1-6 \times
10^5$ years) class of cooling neutron stars.

For \PSRB, which is the oldest among the three pulsars investigated, any hint
of emission from the cooling stellar surface has faded to below what might be
detected in the XMM-Newton observation. The $3\sigma$ surface temperature
upper limit of $T_s^\infty < 480\,000$ K is well above temperatures predicted
by current models of neutron star thermal evolution even if strong frictional
heating of superfluid ${}^1\mbox{S}_0$-neutrons in the outer neutron star crust
is considered (Umeda et al.~1993; Yakovlev et al.~2002).
The same is true for \PSRBB\, where $T_s^\infty < 500\,000$ K and for \PSRJ\,
for which $T_s^\infty < 627\,000$ K. However, standard neutron star cooling
models neglect the influence of a strong magnetic field on the neutron
star's thermal evolution.  As the heat transport in neutron stars is mainly
due to electrons the presence of a magnetic field is supposed to reduce the
thermal conductivity perpendicular to the magnetic field direction. The
consequences are an anisotropic temperature distribution on the neutron star
surface (Geppert, K\"uker \& Page 2004) and a reduced cooling rate so that
magnetic cooling curves may deviate significantly from the zero-field case
after $10^{5-6}$ years (Tsuruta 1998). The surface temperature upper limits
of the old, non-recycled, pulsars, even if they are above what standard cooling
models predict, may still provide interesting constraints for those thermal
evolution models which take the neutron star's magnetic field into account.

As far as emission from a thermal polar cap is concerned, it is very
interesting that there is no clear evidence for the presence of this component
in any of the energy spectra. Yet, emission from a heated polar cap is present
in the spectra from both the younger cooling neutron stars and the older, recycled,
millisecond pulsars (cf.~Becker \& Pavlov 2001) as long as the neutron star is
active as a pulsar. Harding \& Muslimov (2001; 2002) predicted in the framework 
of their revised space-charge-limited flow model that polar cap heating, as a 
fraction of the spin-down luminosity, increases with pulsar age and should be most 
efficient for pulsars of spin-down age {\Large $\tau$}$\sim 10^7$ yrs, if 
they are in fact producing pairs from curvature radiation photons. According to 
these models, however, B0950+08 and B0823+26 cannot produce pairs from curvature 
radiation (CR) of primary electrons since they both lie below the CR pair death 
line in the $P$-$\dot{P}$ diagram of radio pulsars (i.e.~the primary electrons 
cannot accelerate to the energies required to produce CR pairs). Both of these 
pulsars can however produce pairs from inverse Compton scattered (ICS) photons, 
which provide much lower PC heating than do CR-produced positrons resulting in 
predicted luminosities of $L_{+}^{ICS} \simeq 10^{28}\,\rm erg\,s^{-1}$ for 
B0950+08 and $L_{+}^{ICS} \simeq 6 \times 10^{27}\, \rm erg\,s^{-1}$ for B0823+26.  
Both of these values are well below the luminosities that we have observed for 
these sources, and also below the upper limits for emission from a heated polar 
cap which are $L_{pc} < 8.4 \times 10^{28}\,\rm erg\,s^{-1}$ for B0950+08 and 
$L_{pc} < 1.3 \times 10^{29}\,\rm erg\,s^{-1}$ for B0823+26. The results are 
thus consistent with a non-thermal, not a polar cap heating, origin for  
the emission from these two pulsars.  J0243+2740 lies well above the CR pair 
death line of Harding \& Muslimov (2002), and thus is expected to have a much 
higher level of polar cap heating from CR produced positrons with a luminosity 
predicted to be $L_{+}^{CR} \simeq 10^{31}\,\rm erg\,s^{-1}$.  This value is 
near but below our observed luminosity of $L_{obs} =2.8 \times 10^{31}\,\rm 
erg\,s^{-1}$, implying that a significant part of the observed luminosity 
could come from polar cap heating, depending on how  directly we are viewing 
the polar cap. Thus the luminosities detected from all three pulsars are 
consistent with the predicted level of polar cap heating.

\begin{figure}
\centerline{\psfig{figure=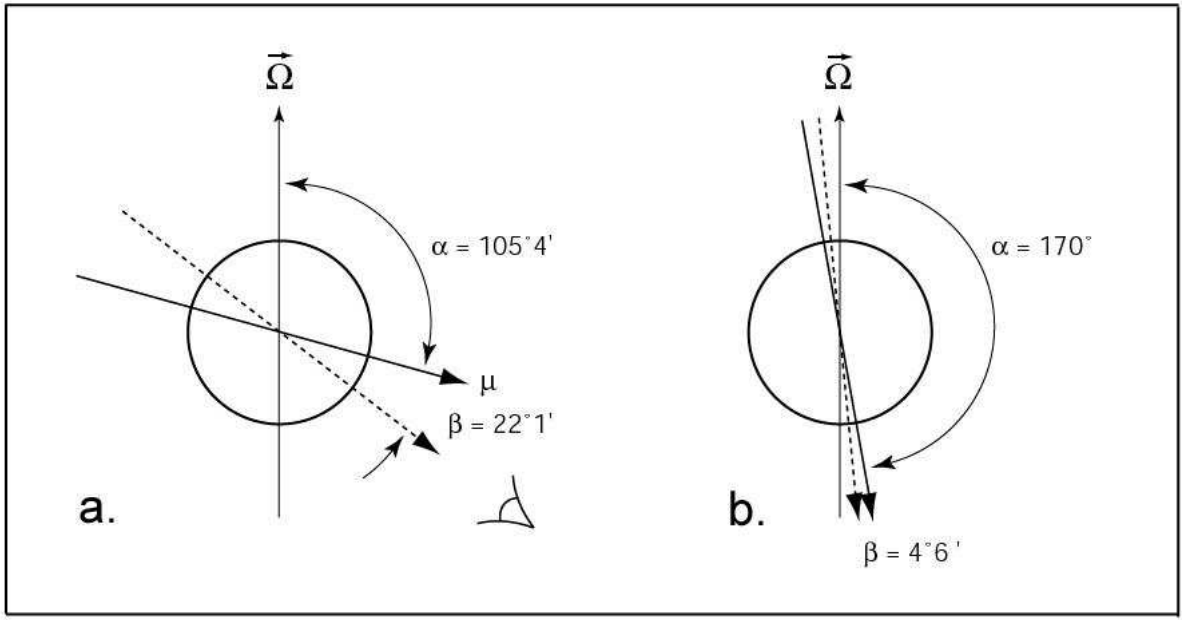,width=16cm,clip=}}
\caption[]{Emission beam geometry of \PSRB\,  for a nearly orthogonal (a.) and almost 
aligned rotator (b.) model. $\alpha$ is the inclination of the magnetic axis, $\beta$
the minimum angle between the magnetic axis and the line of sight. $\vec{\Omega}$ is 
the rotation axis}
\label{PSRB_geometry}
\end{figure}

The geometry of \PSRB\, and \PSRBB\, has been investigated recently by
fitting the classical rotating vector model to high-quality polarization
data taken with Arecibo Observatory at 1.4 GHz (Everett \& Weissberg 2001).
These authors favor the interpretation that both pulsars are almost orthogonal
rotators, but Narayan \& Vivekanand (1982), Lyne \& Manchester (1988), Blaskiewicz,
Cordes \& Wassermann (1991), Rankin (1993a; 1993b) and von Hoensbroech \& Xilouris
(1997) in previous observations came to the conclusion that the emission geometry of
\PSRB\, is that of an almost aligned rotator.
Figure \ref{PSRB_geometry} shows the geometry of the two scenarios for \PSRB\,
with inclination and impact angles taken from Everett \& Weissberg (2001) and
references therein.

Although rotating vector model fits are easily perturbed by systematic effects in 
polarized position angles and reported uncertainties often underestimated the actual 
errors, we find that the double peaked X-ray pulse profile of \PSRB\, with the peak 
separation $\delp \simeq 0.4$ strongly supports the nearly orthogonal rotator.
In the aligned rotator geometry, with an inclination angle of $\sim 170^\circ$
and impact angle of $\sim 5^\circ$ (see Figure \ref{PSRB_geometry}a), the polar 
cap model (Ruderman \& Sutherland 1975; Daugherty \& Harding 1982) predicts a 
single-peaked profile. Double peaked profiles (as those in Daugherty \& Harding 
1996, and Dyks \& Rudak 2002) can be observed only when our line of sight crosses 
the polar gap. This would require the gap to be located at least 15 stellar radii 
above the surface, or the surface conal beam would have to be 4 times wider than
the polar cap beam. Moreover, an improbably fine tuning of model parameters would 
be required to reproduce the large peak separation.

In the outer magnetosphere scenarios such as the outer gap model (Cheng,
Ho \& Ruderman 1986; Romani \& Yadigaroglu 1995; Cheng, Ruderman \& Zhang 2000) 
or the two-pole caustic model (Dyks \& Rudak 2003), the non-thermal X-rays are 
emitted in a fan beam. For the nearly aligned geometry, however, the outer gap 
model predicts no high-energy radiation (see top panel in Figure 6 in Cheng, Ruderman 
\& Zhang 2000) whereas the two-pole caustic model predicts single-peaked 
lightcurves (see Figure 2b in Dyks \& Rudak (2003).

\begin{figure}
\centerline{\psfig{figure=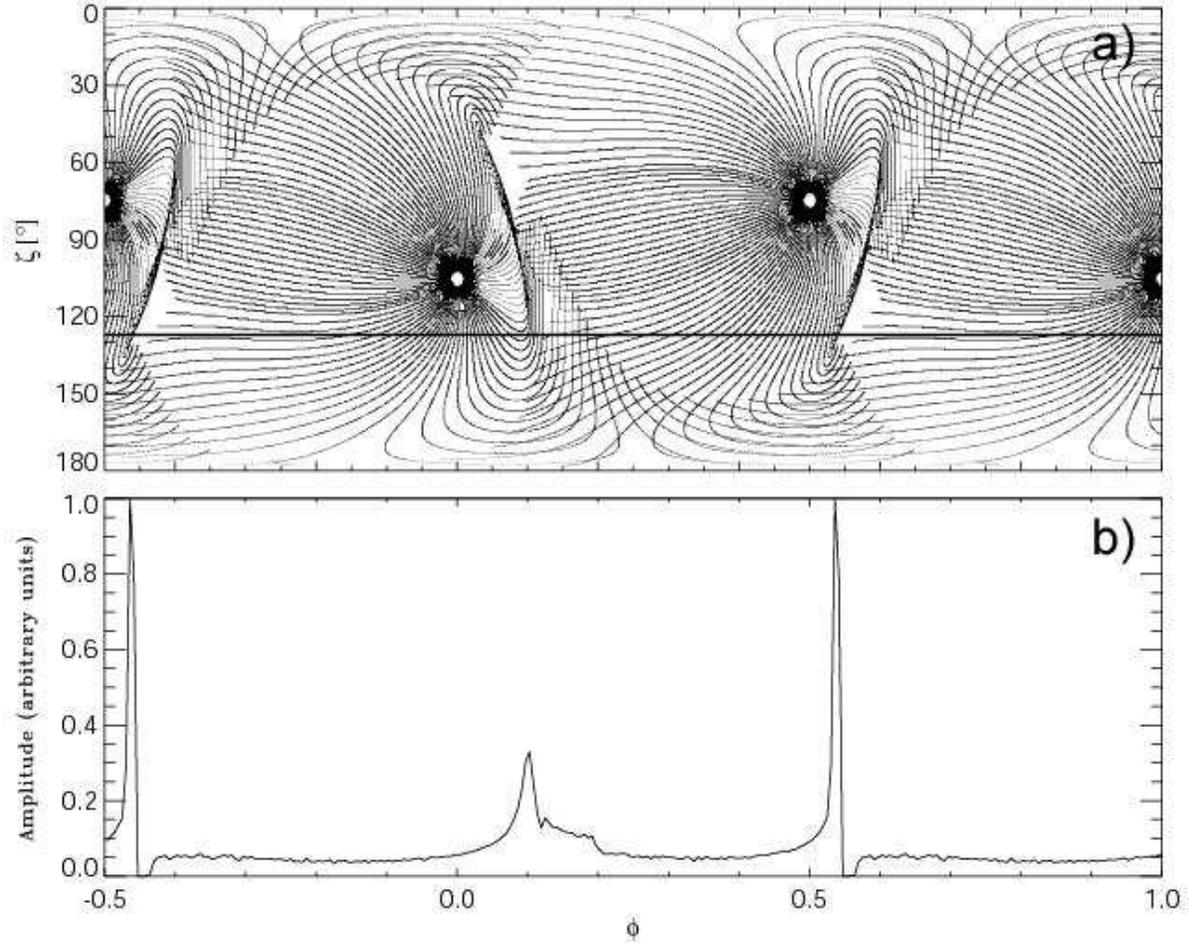,width=16cm,clip=}}
\caption[]{The radiation pattern (a.) and the high-energy pulse
profile (b.) obtained with the two-pole caustic model
for \PSRB\ with $\alpha = 105^\circ.4$ and $\zeta=127^\circ.5$.
The profile corresponds to the horizontal cut through the
radiation pattern marked with the straight line in {a.}.
Each peak is associated with a different magnetic pole.
The flux is in arbitrary units and one and a half periods are shown.
}
\label{tpc}
\end{figure}

Both the outer gap and two-pole caustic models can reproduce the observed profile 
in the nearly orthogonal scenario. Figure \ref{tpc} presents the radiation pattern 
({\it top}) and the pulse profile ({\it bottom}) calculated for the two-pole caustic 
model with the dipole inclination $\alpha = 105^\circ$ and and the viewing angle 
$\alpha+\beta = 127^\circ \equiv \zeta$, as derived by Everett \& Weisberg (2000).
Each peak arises due to the caustic effects on the trailing side of the open field 
line region associated with each magnetic pole (see Figure \ref{tpc}a). The modeled 
peak separation ($\delp \simeq 0.43$) is in good agreement with the observed one 
($\sim 0.4$). The {\it relative} widths and heights of these two peaks also resemble 
the observed ones, however, they are more model-dependent than $\delp$.

Our three-dimensional simulations show that for the parameters given above, the outer
gap model also predicts a double-peaked profile, with $\delp \le 0.3$ which is
marginally consistent with the data. The outer gap model can more closely reproduce 
the large peak separation for viewing angles closer to the rotational equator 
($\zeta \sim 100^\circ$). However, according to outer gap models, only younger 
pulsars can sustain a gap in their magnetospheres and produce non-thermal 
high-energy emission. B0950+08, B0823+26 and J0243+2740 all lie below the 
original outer gap death lines for production  of high-energy emission 
(Chen \& Ruderman 1993), indicating that they do not produce high-energy emission 
from outer gaps.  More recently, the outer gap death lines have been revised to 
include pulsar inclination and thermal emission from cooling and heated polar caps 
(Zhang et al.~2004). All three pulsars lie above at least one of the revised outer gap 
death lines computed by Zhang et al.~(2004), so that outer gap emission may not be ruled 
out.

Neither the two-pole caustic nor the outer gap model can explain the relative locations
of the X-ray and radio peaks. Both models predict that the main (i.e., the strongest)
radio peak should precede the leading X-ray peak in phase roughly by $\sim 0.1$. This 
is the phase at which our line of sight approaches most closely one of the magnetic 
poles (Figure \ref{tpc}a). The standard polar cap model in the nearly orthogonal geometry
cannot explain the X-ray lightcurve. Because the closest approach to a magnetic pole 
occurs near the leading peak, the model predicts that this peak should be stronger 
and more spiky than the trailing peak.

In the case of \PSRBB\, multiple emission components are seen in the radio 
pulse profile while the statistics of the available XMM-Newton data is not 
sufficient to better resolve the X-ray pulse profile than to a single broad 
peak.

Non-thermal X-radiation processes implied by the outer magnetosphere 
interpretation are in agreement with the non-thermal spectra which dominate 
the emission from all three pulsars. This is most evident in the energy spectrum
of \PSRB\, but appears to be the case for both \PSRBB\, and \PSRJ\, as well.
However, the predicted level of polar cap heating for \PSRJ\ implies that a significant 
part of its emission may be thermal. 

The optical emission from \PSRB\, has been recently observed with the VLT FORS1
(Zharikov et al.~2003). Fitting the optical data simultaneously with the X-ray 
spectrum suggests a broadband spectrum which can be described by a broken power 
law, strongly suggesting that the radiation from \PSRB\, is dominated by
non-thermal emission from the optical to the X-ray band. Even more, taking \PSRB,
\PSRBB\, and \PSRJ\, as representative would imply that the X-ray emission from old, 
non-recycled, rotation-driven pulsars is dominated by non-thermal radiation as has 
been concluded by Becker \& Tr\"umper (1997) based on their tight soft-X-ray luminosity 
vs.~spin-down energy conversion fits of ROSAT detected rotation-powered pulsars. The emission 
properties observed from PSR B1929+10 (cf.~\S \ref{intro}) are not in disagreement 
with this conclusion. XMM-Newton observations of this pulsar, which have taken place in
November 2003 and April 2004, are expected to further constrain this conclusion.

\acknowledgments
AJ would like to thank M.~Kramer for the supply of new pulsar ephemeris data
and M.~Kramer and O.~Doroshenko for advice on the use of the TEMPO and TIMAPR
pulsar timing packages. J.D. acknowledges a National Research Council Research 
Associateship Award at NASA/GSFC. We thank the anonymous referee for thoroughly reading the manuscript and the many useful comments.

\begin{deluxetable}{lccc}
\tablewidth{0pc}
\tablecaption{Radio Properties
\label{t:radio}}
\tablehead{}
\startdata
	                                  & \PSRB & \PSRBB & \PSRJ \\ \hline
Period (ms)                               & 253   & 530	   &  96   \\
Spin-down age (yr/$10^6$)                 & 17.37 & 4.89   & 1.2   \\
Spin-down energy ($\mbox{erg/s}/10^{32}$) & 5.62  & 4.57   & 562   \\
Inferred Magnetic Field ($G/10^{11}$)     & 2.45  & 9.77   & 3.54  \\
Dispersion Measure ($\mbox{pc/cm}^3$)     & 2.97  & 19.47  & 21    \\
Distance$^a$ (pc)                         & 255   & 340    & 1130  \\
$n_e$ ($\mbox{cm}^{-3}/10^{2})$           & $1.13^b\pm 0.02$ & $\sim 5.72$$^c$ & $\sim 1.86$$^c$ \\
$N_H$ ($\mbox{cm}^{-2}/10^{19}$)          & 9.6 & 60 & 65 \\
\enddata
\tablecomments{\newline
$^a$ Dispersion-measure inferred distance (Cordes and Lazio 2002). \\ 
$^b$ Brisken et al.~(2002). \\
$^c$ Computed using $DM/d$ and the electron density model NE2001 of Cordes \& Lazio (2002).
}
\end{deluxetable}

\begin{deluxetable}{llccccc}
\tablewidth{0pc}
\tablecaption{Instrument setups, filter usage, start time and
durations, and effective exposures
of the XMM-Newton observations of \PSRB, \PSRBB\, and
\PSRJ.\label{observations}}
\tablehead{}
\startdata
Detector & \quad\quad\quad Mode & Filter & Start time
& Duration & eff.~Exp. \\
{} & {} & {} & (UTC)
& ksec & ksec \\\hline\\[-1ex]

\multicolumn{5}{c}{PSR B0950+08}\\\\[-2ex]
EMOS1 & PrimeFullWindow & Medium & 2002-05-08T18:13:31 & 13.9 & 8.8\\
EMOS1 & FastUncompressed& Medium & 2002-05-08T22:14:31 & 27.5 & {} \\
EMOS1 & PrimeFullWindow & Medium & 2002-05-09T06:06:25 & 13.9 & 11.0\\
EMOS1 & FastUncompressed& Medium & 2002-05-09T10:07:26 & 25.7 & {} \\
EMOS2 & PrimeFullWindow & Medium & 2002-05-08T18:13:32 & 83.7 & 65.7 \\
EPN & PrimeSmallWindow& Medium & 2002-05-08T18:29:20 & 82.9 & 48.9\\\hline\\[-1ex]
\multicolumn{5}{c}{PSR B0823+26}\\\\[-2ex]
EMOS1 & PrimeFullWindow & Medium & 2002-04-26T17:43:18 & 18.5 & 17.9\\
EMOS1 & FastUncompressed & Medium & 2002-04-26T23:00:11 & 31.3 & {} \\
EMOS2 & PrimeFullWindow & Medium & 2002-04-26T17:43:16 & 50.7 & 47.7 \\
EPN & PrimeSmallWindow & Medium & 2002-04-26T17:59:03 & 49.9 & 33.9 \\ \hline\\[-1ex]
\multicolumn{5}{c}{PSR J2043+27}\\\\[-2ex]
EMOS1 & PrimeFullWindow & Medium & 2002-11-21T23:21:42 & 16.7 & 16.5 \\
EMOS2 & PrimeFullWindow & Medium & 2002-11-21T23:21:48 & 16.7 & 16.5 \\
EPN & PrimeSmallWindow & Thin & 2002-11-21T23:26:20 & 16.5 & 10.8 \\ \hline
\enddata
\end{deluxetable}


\begin{deluxetable}{cccccc}
\tablewidth{0pc}
\tablecaption{Spectral Fits for \PSRJ
\label{spectral_fits_J2043}}
\tablehead{}
\startdata
model$^a$ & $\chi_\nu^2$ & $\nu$ & $N_H/10^{22}$ &$\alpha$ / $kT$$^b$ & Radius$^c$ \\
{} & {} & {} & $\mbox{cm}^{-2}$ & {} & km \\\hline\\[-1ex]

bb & 1.12 & 7 & $0.0^{+0.3}$ & $0.14_{-0.06}^{+0.03}$ & $0.4_{-0.2}^{+0.2}$ \\\\[-1ex]

bb & 1.37 & 8 & $0.4_{-0.2}^{+0.2}$ & $0.08_{-0.01}^{+0.01}$ & 10 \\\\[-1ex]

pl & 0.53 & 7 & $0.0^{+0.2}$ & $3.1^{+1.1}_{-0.60}$ & {} \\\\[-1ex]

pl+bb & 0.67 & 7 & 0.065 & $2.8_{-0.8}^{+1}/ < 0.125$ & 0.467 \\\\[-1ex]

pl+bb & 0.49 & 7 & 0.065 & $2.7_{-0.9}^{+1.3}/ < 0.054$ & 10 \\\\[-1ex] \hline

\enddata
\tablecomments{
$^a$ bb = blackbody; pl = power law; \\
$^b$ The entry in this column depends on the spectral model --- it is the
power law photon index $\alpha$ or the temperature $kT$ in keV\\
$^c$ For models for which we fixed the radius of the emitting area we
assumed
a pulsar distance of 1.8 kpc.
}
\end{deluxetable}

\clearpage

\clearpage

\clearpage

\clearpage

\clearpage

\clearpage

\clearpage

\clearpage

\clearpage

\clearpage

\clearpage

\clearpage

\clearpage

\end{document}